\documentclass[intlimits,twoside,a4paper]{article}

\usepackage{amsmath,amssymb}   
\usepackage[T2A]{fontenc}
\usepackage[cp1251]{inputenc}

\usepackage{color}

\usepackage{cmpj2}

\issue{2014}{17}{1}{12601}
\doinumber{10.5488/CMP.17.12601}

\articletype{Review Article}

\title[Fluid-grafted chains interfaces]
{Description of interfaces of fluid-tethered chains: advances in density functional theories and off-lattice computer simulations}

\author[S. Soko\l owski, J. Ilnytskyi, O. Pizio]{
S. Soko{\l}owski\refaddr{label1},
J. Ilnytskyi\refaddr{label2},
O. Pizio\refaddr{label3}\thanks{E-mail: oapizio@gmail.com}}

\addresses{
\addr{label1} Department for the Modelling of Physico-Chemical Processes,
Maria Curie-Sklodowska University, 33~Gliniana~St., Lublin, Poland
\addr{label2} Institute for Condensed Matter Physics of the National Academy of Sciences of Ukraine, \\ 1~Svientsitskii St., 79011 Lviv, Ukraine
\addr{label3} Instituto de Qu\'{i}mica, Universidad Nacional Autonoma de M\'{e}xico,
Circuito Exterior, Ciudad Universitaria, 04510 M\'{e}xico, D.F., M\'{e}xico}

\date{Received May 28, 2013, in final form October 24, 2013}
\authorcopyright{S. Soko\l owski, J. Ilnytskyi, O. Pizio, 2014}

\begin{document}

\maketitle

\begin{abstract}
Many objects of nanoscopic
dimensions involve fluid-tethered chain interfaces.
These systems are of interest for basic science and for several applications, in
particular for design of nanodevices for specific purposes.
We review recent developments of theoretical methods in this area of research
and in particular of density functional (DF) approaches, which provide
important insights into microscopic properties of such interfaces.
The theories permit to describe the dependence of adsorption, wettability,
solvation forces and electric interfacial phenomena on thermodynamic
states and on characteristics of tethered chains.
Computer simulations for the problems in question are overviewed as well.
Theoretical results are discussed in relation to simulation results
and to some experimental observations.
\pacs  61.20.-p, 61.25.-f, 61.25.he, 61.25.Em, 68.15.+e, 68.43.-h, 68.47.Pe, 78.30.cd
\end{abstract}

\section{Introduction}

Systems involving interfaces formed by polymers tethered onto a solid surface and a fluid
contacting them are widespread in nature and are the subject of many studies.
One of the reasons is that the grafted  layers of chains find  several
practical applications, for example in efficient colloid stabilization, protection
and prevention of surfaces from undesirable adsorptions, manufacturing
of functional surfaces and nanoscale fluidic channels, and many others.
Interesting properties of such systems have also stimulated the development of several
methods of synthesis  for preparation of assemblies of
end-grafted polymers.  However, up to now, the
development of new preparation methods is mainly based on empirical
knowledge or intuition.

From the very beginning, experimental studies of tethered chains are
accompanied by the development of theoretical methods of their description.
They involve thermodynamic and structural properties of systems
containing uncharged and charged species under the effect
of various external factors including confinement at nanoscale as well.

The purpose of this article is to present a brief review of the
 statistical mechanical DF methods
and off-lattice computer simulations of classical
fluids in contact with the walls modified by tethered polymers,
to discuss some recent achievements in this area of research, as well as
to provide some prospects for further investigations.
After introductory remarks, in the next section (section~\ref{sec2}), we
briefly outline different types of theoretical approaches
used for the description of tethered chains-fluid interfaces.
We begin with the scaling theories, but our emphasis
is on the theories that are based on the microscopic description
and the use of the intermolecular potentials as the input, i.e.,
on the integral equations, self-consistent field and
density functional theories. Particular attention is paid
to the  DF approaches.
In section~\ref{sec3}
we discuss one version of the density functional theory (DFT) that
has been successful in describing several phenomena occurring
within tethered chains-fluid interfaces. Some illustrative examples
are given in section~\ref{sec4}. Section~\ref{sec5} extends the approach
of section~\ref{sec4} to the case of systems having electrostatic interactions.
We discuss the theory and then, in section~\ref{sec6},
present some applications concerning structural, thermodynamic and
electric properties of the systems with tethered chains.
Mesoscopic simulations by the dissipative particle dynamics method
and the (semi-) atomistic simulations using the
Monte Carlo (MC) and Molecular Dynamics (MD) techniques are reviewed
in sections~\ref{sec7a} and \ref{sec7b}, respectively.
The manuscript ends with a short summary in section~\ref{sec8}.

\section{Overview of theoretical methods}\label{sec2}

Essential features of the systems involving
end-tethered polymers onto a solid and in contact with a fluid
can be captured by modelling at atomistic or semi-atomistic
(at groups of atoms) scale by using contemporary theories for interfaces
between a solid and a complex fluid.
A link between fundamentals and applications
can be reached by combining theoretical (classical and quantum)
tools, intensive computer simulations and laboratory experiments.
Current studies at a simple level of modelling
put focus on the development of a theory for various
phenomena occurring in the systems of our interest.
Computer simulations
 provide data to confront theoretical predictions and simulation results.
The most important, they serve as predictive tools
to guide elaboration of interfaces with desired functional properties.

Theoretical tools to investigate systems with
tethered chains  are similar to the methods
designed to study nonuniform polymeric systems.
They can be classified into several groups by
using different criteria. One can divide them
according to the way in which translational degrees of
freedom are considered. Namely, either
lattice or off-lattice models can be distinguished.
On the other hand, the classification can
be made according to  theoretical approaches used
to describe a model.
One can distinguish the theories based on the application
of scaling arguments \cite{gennes,gennes1},
integral equations approaches, self-consistent field (SCF)
as well as the DF methods.
The scaling approaches provide qualitatively correct
trends of the system behavior over a variety
of conditions. The integral equations and the SCF approaches
are  more ample in the sense that they provide information (e.g., details
on the structure of interfaces) which scaling approaches do not.
Perhaps, the most accurate are the DFTs that besides an insight into the structure
also provide an elegant way to study phase transitions in the systems.
Finally, one can classify the models into the ones with or
without electrostatic interactions.
The systems involving tethered molecules have been also
extensively studied using computer simulation methods.

\subsection{Scaling approaches. A few comments}

A system to which the below described theoretical methods apply, is
a set of long flexible polymer chains which are anchored on a grafting
substrate through  a special single end group.
In the majority of cases
in question, the substrate is taken as flat.  The degree of polymerization determines
the ``chain length'' $M$. On the other hand, the important parameter of the system is
the grafting density $\rho_\textrm{C}$, i.e., the fraction of grafted sites per unit
area. Usually,  this system  is put in contact with a solvent.
These three parameters ($M$, $\rho_\textrm{C}$ and solvent quality) determine
several regimes of behavior of polymer layer having distinct static
(structural and thermodynamic) and dynamic properties. The systems of this type
are intrinsically characterized by a set of mesoscopic length scales which
makes the polymer layers under different conditions amenable to
the scaling analysis.
Of course,  scaling relation imply that all characteristic lenghts are
much larger than any atomistic scales, e.g., the range over which
the grafting substrate induces density oscillations or the persistence length.

It is commonly assumed that all other segments of chain molecules,
besides grafted end group, are not  attracted to the surface.
Within the scaling theories, the structure of a grafted
polymer layer is characterized by the average height  and
the values of gyration radius along selected axis of grafted molecules.
If the grafting density is very low, different coils do not overlap and
practically do not interact. Each chain occupies roughly a
half-sphere with a radius comparable to the Flory radius for a polymer
coil in a good solvent, such that values of gyration radius along selected axis in the $z$-direction normal to
the surface $R_{\textrm{g}z}$  and in the directions perpendicular to
the $z$-axis $R_{\textrm{g}\perp}$  differ only by their prefactor in the
power law relating size to the chain length~\cite{gennes,gennes1},
\begin{equation}
 R_{\textrm{g}z}\propto R_{\textrm{g}\perp}\propto M^\nu
\end{equation}
with $\nu\approx 3/5$,  $M$ is the
number of segments forming a chain.
This behavior is the so-called ``mushroom regime''.

For denser grafted layers,  the chains overlap close to
the grafting surface, and are bound
to stretch away from the surface, like
the bristles in a brush. This situation is quite different
from a typical behavior of polymer chains in a solution.
If the grafting density $\rho_\textrm{C}$ is high (the polymer brush regime)
several  characteristic length scales appear.
According to the scaling approach by Alexander \cite{alexander}
 each grafted polymer chain is considered as a linear array
 of $M'$ ``blobs'', $M'=M/n_\textrm{blob}$, where $n_\textrm{blob}$
 is the number of chain segments per blob (only segements of a single
chain are inside a blob),
 $n_\textrm{blob}\propto \xi_\textrm{g}^{1/\nu}$  ($\xi_\textrm{g}=\rho_\textrm{C}^{-1/2}$
 is the average distance between the grafting points or
 the blob diameter).

 The theory by Alexander, see reference~\cite{alexander}  for details,
yields the following
 scaling law for the tethered polymer layer height, $h$,
 \begin{equation}
  h\propto \rho_\textrm{C}^{(1/\nu-1)/2}M\approx \rho_\textrm{C}^{1/3}M.
 \end{equation}
This is a very important result showing that the brush height
at a chosen fixed grafted density,
is proportional to $M$, while the unstretched chain dimension, $R_\textrm{g}$,
grows slower.  Consequently, the properties of chains under
such conditions may be expected  to differ much from unstretched
chains in a solution. This scaling argument implies
a step-like density profile of the segments.
It is difficult to expect a such situation for more realistic models.

In fact, entropy favors a nonuniform blob
 picture. A nonuniform stretching of chains can occur, some
 chains can turn backwards and some segments
 can be located anywhere within the brush (of course,
 there is a maximum distance from the surface within
 which the segments can be found). Due to nonuniformity,
 with an increasing distance from the
 surface,  $\xi_\textrm{g}$ grows from the value $\rho_\textrm{C}^{-1/2}$ assumed by
Alexander, to larger values.

In particular, for semi-dilute grafted layers, grafted at $z=0$,
as documented by de Gennes~\cite{gennes1}, the density profile
$\xi_\textrm{g}(z)=\xi_\textrm{g}(0)[\rho_\textrm{s}(z)/\rho_\textrm{s}(0)]^{[\nu/(1-3\nu)]}$,
is related to the segment local density $\rho_\textrm{s}(z)$,  that
is assumed to be parabolic, $\rho_\textrm{s}(z)=\rho_\textrm{s}(0)[1-(z/h)^2]$
in the self-consistent field theory commented below.
The scaling approaches do not take into account the
interactions in the system explicitly.
Consequently, specific correlations  between particular species of importance
in various phenomena cannot be captured in such framework.

\subsection{Integral equations}

Development of theoretical approaches in which intermolecular potentials
are the input to the theory is particularly important for
discussing phase transitions.
One of the first theoretical approaches to describe uniform,
as well as nonuniform polymeric systems that explicitly
takes into account interaction potentials between all
the species was based on the  integral equation
theory of fluids. Common integral equation theories
of atomic fluids
were extended to describe polymers by
introducing the so-called
Polymer Reference Interaction Site Model (PRISM)~\cite{schweizer,heine,hall}
that generalized the Reference Interaction Site Model (RISM)
by Chandler~\cite{chandler0}.
For the case of a fluid of homopolymers
built of $M$ indentical sites (segments), the PRISM integral equation
in the Fourier space  reads,
\begin{equation}
 \tilde{h}(k) = \tilde\omega(k) \tilde c (k) [\tilde \omega (k)+
\rho_M \tilde h(k)],
\label{PRISM1}
\end{equation}
where $\rho_M$ is the site density ($\rho_M=M\rho$, with $\rho$ being the density of molecules),
$ \tilde h(k)$ is the Fourier transform of the site-site total correlation function, $ \tilde c(k)$
is the Fourier transform of the site-site direct correlation function. Since all the sites along
a chain molecule
are equivalent, it follows that  $\tilde h(k) = \tilde h_{\alpha,\gamma}(k)$ and
 $\tilde c(k) = \tilde c_{\alpha,\gamma}(k)$.
Moreover,
\begin{equation}
 \tilde \omega(k)=\frac{1}{M}\sum_{\alpha,\gamma=1}^M
\tilde  \omega_{\alpha,\gamma}(k),
\label{PRISM2}
\end{equation}
is the single chain structure factor consisting of Fourier transforms of the site-site
intramolecular correlation functions.
$\omega_{\alpha,\gamma}(k)$. A generalization of equations~(\ref{PRISM1})--(\ref{PRISM2})
to the cases of mixtures of molecules with different types of  sites is available
in the literature \cite{schweizer}.
After introducing  closure approximations, the system
of the PRISM integral equations can be solved numerically,
yielding  the correlation functions that contain a
complete information about the microscopic structure of the
molecular system.
By computing the integrals of the correlation functions,
one can obtain thermodynamic properties.

The PRISM-based approaches have been used to study
a type of model systems involving  small
solid particles of different geometry (e.g., spherical),
coated with polymer chains, see e.g., closely related
works \cite{hall,gast,hasegawa,semenov}.
One recent interesting example is due to Jayaraman and
Schweizer~\cite{jayaraman1,jayaraman2,jayaraman3}
who used the PRISM theory to study model polymer nanocomposites
consisting of nanoparticles (called ``fillers'') with a few grafted
chains embedded in a polymer ``matrix'' (solvent or melt).
Coated nanoparticles with a relatively small and controlled number
of  grafted polymers in this type of mixtures are of interest for various applications.
Both the microscopic structure and the phase behavior  of such complex
systems are  sensitive to the effective interactions between species.
Those can be tuned by chemical means: synthetic routes
of laboratory preparation determine the attraction strength
between bare nanoparticles, tether grafting density and matrix chain length.
Other important factors are the total packing density and mixture composition,
along with the diameter ratio between nanoparticles and polymerizing monomers
belonging to grafted chains and to the polymer solvent.
We would  just like to mention
three scenarios leading to  the development of different structures.

In general terms, there is a competition between the assumed
nanoparticle-nanoparticle attraction, steric repulsion,
i.e., excluded volume effects between grafted tethers, and
filler-homopolymer correlations.
The matrix subsystem can induce a depletion attraction
between fillers.
Under certain circumstances, the
tether-tether repulsion can sterically ``screen'' the attraction
between nanoparticles and if the depletion type entropic attraction
is weak, then the fillers form a well-dispersed fluid of a filler in
the homopolymer matrix. If the intertether repulsion is weak and
the attraction between fillers is strong enough, then the macroscopic
phase separation can be induced by depletion matrix effects in addition
to bare attraction between nanoparticles. In this case, the matrix-rich
and the tethered filler-rich coexisting phases can be formed.
Still another, ``intermediate''  physical situation occurs if nanoparticle-nanoparticle
interactions are rather strong and a sufficient steric stabilization
due to tether repulsion exists to preclude macrophase demixing,
then aggregate formation of microphase-type ordering can take place.
So far the PRISM theories have not been applied to describe
situations where the grafting occurs on planar surfaces, or, in
general, on the surfaces with linear dimension much larger than the size
of a polymer segment.

\subsection{Self-consistent field approaches}

One of the popular approaches
for the systems of our interest is the self-consistent field theory (SCFT),
developed by
Edwards~\cite{edwards}, DiMarzio~\cite{dimarzio}, Vrij~\cite{vrij} and
Helfand and Tagami~\cite{helfand1,helfand2}.
The key idea behind SCFT is to approximate an ensemble of interacting polymers
by the system of mutually non-interacting polymer
molecules in an effective, position dependent field.
The SCFT methods have been successful in predicting
adsorption of polymers, tethered polymers,
polymer blends and copolymers,
but they are intrinsically incapable of capturing fine details of the system structure
at short-range scale~\cite{schmidf,matsen,fleer}.
Due to space limitations, we would like to restrict the discussion of the method
to the statement of the problem rather than to the enumeration of  several principal results.

For illustrative purposes, we begin with a general statistical mechanical arguments and
consider the Gaussian chain model.  The system is a blend of $N_\lambda$  kinds of
linear homopolymers.
The length of the homopolymers of a given kind $\lambda$ is $M_\lambda$.
For Gaussian chains, the potential energy of the system is \cite{freed0}
\begin{equation}
 \beta V_\textrm{pot}= \sum_{\lambda\alpha} \sum_{i=1}^{{M_\lambda}-1} \frac {3}{2l_{\lambda}^2}
\left|{\bf r}_{\alpha,i+1}^\lambda - {\bf r}_{\alpha,i}^\lambda\right|^2 +
\frac{\beta}{2} \sum_{\lambda\alpha i \ne \lambda'\alpha' i'}
u_{\lambda\lambda'} \left({\bf r}_{\alpha,i}^\lambda - {\bf r}_{\alpha',i'}^{\lambda'}\right)+
\beta \sum _{\alpha\lambda i} v_{\lambda}\left({\bf r}_{\alpha,i}^\lambda\right),
\label{eq:vpot}
\end{equation}
the sum over $\alpha\lambda i$ is over all the species, over all the particles of the given species and
over all the sites of a molecule of a given kind and $l_\lambda$ is the corresponding Kuhn length;
$\beta=1/kT$,
$u_{\lambda\lambda'}$  and  $v_{\lambda}$ are the segment-segment interaction potential and
the external field, respectively.
The grand partition function for the model in question is,
\begin{equation}
\Xi = \prod_{\lambda} \sum_{{n_\lambda}=0}^{\infty} \exp\left(\beta\sum_{\lambda}
\mu_{\lambda}n_{\lambda}\right)
\frac {\left(3/2\pi l_{\lambda}^2\right)^{3(M_\lambda-1)/2} }
{\prod_{\lambda}
 \left[2^{n_\lambda} (n_\lambda ! )
\Lambda_{\lambda}^{3{n_\lambda}{M_\lambda}}\right]} \left\{ \prod_{\lambda}\prod_{\alpha=1}^{n_\lambda} \prod_{i=1}^{M_{\lambda}}
\int \rd{\bf r}^{\lambda}_{\alpha,i} \right\} \exp\left(-\beta V_\textrm{pot}\right),
\label{eq:freed1}
\end{equation}
where $\mu_\lambda$ is the chemical potential per chain of species $\lambda$, and
$\Lambda_\lambda = h/(2\pi m_\lambda kT)^{1/2}$ is the thermal de~Broglie length for monomers
of species $\lambda$.

It is convenient to define the effective field function $W_{\lambda}({\bf r})$  via the relation
\begin{equation}
 \int \rd {\bf r} \sum_\lambda \tilde \rho_\lambda ({\bf r}) W_\lambda({\bf r}) =
\beta \sum_\lambda \mu_\lambda n_\lambda -\beta \sum_{\lambda\alpha i} v \left({\bf r}_{\alpha i}^\lambda\right),
\end{equation}
where the density operator for the species $\lambda$ has been introduced
\begin{equation}
 \tilde \rho_\lambda ({\bf r})
 =\sum_{\alpha i}\delta\left({\bf r}-{\bf r}_{\alpha,i}^{\lambda}\right).
\end{equation}
Then, the partition  function (\ref{eq:freed1}) can be rewritten as
\begin{eqnarray}
 \Xi &=& \prod_{\lambda} \sum_{{n_\lambda}=0}^{\infty}
\left[\prod_{\lambda}
 2^{n_\lambda} (n_\lambda !)\Lambda_{\lambda}^{3{n_\lambda}{M_\lambda}}  \right]^{-1}
\left( \frac{3}{2\pi l_{\lambda}^2} \right)^{3(M_\lambda-1)/2} \nonumber\\
&&\times\left\{ \prod_{\lambda}\prod_{\alpha=1}^{n_\lambda} \prod_{i=1}^{M_{\lambda}}
\int \rd{\bf r}^{\lambda}_{\alpha,i} \right\} \exp\left[-\beta V_\textrm{pot}
+ \int \rd {\bf r} \sum_\lambda \tilde \rho_\lambda ({\bf r}) W_\lambda({\bf r})\right].
\label{eq:freed2}
\end{eqnarray}
 The segment density profile is obtained from $\Xi$ by using the definition
\begin{equation}
 \rho_{\lambda}({\bf r}) =\langle\tilde \rho_{\lambda}({\bf r})\rangle  = \frac {\delta \ln \Xi}{\delta W_\lambda ({\bf r})}\,.
\label{eq:freed3}
\end{equation}
Then, the free energy follows from the Legendre transformation \cite{freed0}. In general terms,
the density-field relation (\ref{eq:freed3}) can be inverted analytically for an ideal system to yield
 $\delta W_\lambda ({\bf r})$ as a function of segment density profiles $\rho_\lambda({\bf r})$.
Such a task can be realized for the Gaussian chain ideal model
that permits to reach the principal equations of the self-consistent field method. Tang and Freed \cite{tangfreed}
have chosen the model with a vanishing pair interaction, $u_{\lambda \lambda'} =0$,  as the ideal system
such that the potential energy given by equation~(\ref{eq:vpot}) contains solely the connectivity
of chains contribution as well as the term coming from an arbitrary external field. The partition function
of the model  factorizes as follows:
\begin{equation}
 \Xi_\textrm{ideal}=\prod_\lambda \Xi_\textrm{ideal}^\lambda=\prod_\lambda \sum_{{n_\lambda}=0}^{\infty}
\frac{Q_\lambda^{n_\lambda}}{n_\lambda !}=\prod_\lambda \exp(Q_\lambda),
\end{equation}
where $Q_\lambda$ denotes the partition function for a single Gaussian chain belonging to
the species $\lambda$ under the effect of the arbitrary external field $W_\lambda$.  The
density profile of segments can be easily obtained,
\begin{align}
 \rho_\lambda ({\bf r}) = \frac {\delta \ln \Xi}{\delta W_\lambda ({\bf r})}
= &\frac{\left(3/2\pi l_\lambda ^2\right)^{3(M_\lambda-1)/2}}{2\Lambda_\lambda^{3M_\lambda}}
\left(\prod_{i=1}^{M_\lambda} \int \rd{\bf r}_i^\lambda \right) \left [\sum_i \delta\left({\bf r}-{\bf r}_i^\lambda\right)  \right ]\nonumber\\
&\times
\exp \left [ -\sum_i^{M_\lambda -1}\frac {3}{2l_{\lambda}^2} \left|{\bf r}_{i+1}^\lambda - {\bf r}_{i}^\lambda\right|^2
+\sum_i^{M_\lambda} W_\lambda \left({\bf r}_i^\lambda\right) \right ]
\end{align}
 such that
\begin{equation}
 \int \rd{\bf r} \rho_\lambda (\bf r) = M_\lambda Q_\lambda\,.
\end{equation}

It is more convenient to develop the following  derivation presented below using a continuous chain notation, or,
in other words, substituting sums over monomer indices by integrals. The single chain
partition function, $Q$ (the subscript $\lambda$ is dropped to simplify the notation) reads
\begin{equation}
 Q=\left(1/2\Lambda^{3M}\right) \int \rd{\bf r} \rd{\bf r}' G\left({\bf r},{\bf r}';M,0|[W]\right),
\end{equation}
where $G({\bf r},{\bf r}';M,0|[W])$  is the unnormalized end-vector distribution  function,
given as
\begin{equation}
 G({\bf r},{\bf r}';M,0|[W]) = \int_{{\bf r}(0)={\bf r}'}^{{\bf r}(M)={\bf r}} D[{\bf r}(\tau)]
 \exp \left[ -\frac{3}{2l^2} \int_0^M \rd\tau  |{\bf r}(\tau)|^2  + \int_0^M \rd\tau W[{\bf r}(\tau)] \right].
\label{eq:freed4}
\end{equation}
The chain contour variable $\tau$ changes between $0$ and $M$.  In general, the end-vector
distribution function  $G({\bf r},{\bf r}';\tau,\tau '|[W])$ gives the number of chain
configurations that start at ${\bf r}',\tau '$ and end at ${\bf r}, \tau$.
The path integral representation of the end-vector distribution function given by equation (\ref{eq:freed4})
may be alternatively transformed into the partial differential equation
\begin{equation}
 \left [\frac{\partial}{\partial \tau} - \frac{l^2}{6}\nabla_{{\bf r}}^2 -W({\bf r})\right]
G({\bf r},{\bf r}';\tau,\tau '|[W])=\delta ({\bf r}-{\bf r}')\delta(\tau - \tau ')
\label{eq:freed5}
\end{equation}
that describes a diffusion in an external field $W({\bf r})$.
This type of equations serves  as the starting point in several research works
performed by using the SCFT, see e.g.,~\cite{gast,gast2}. Additional boundary
conditions can be used for a set of specific problems that include
confinement effects.
It is worth mentioning that the inversion of a field-density relation can be explored
in a wider context for systems of interacting chain molecules. The present
simple illustration intrinsically leads us to the explanation of the density functional
approaches.

\subsection{Density functional approaches. Generalities}

The powerful group of methods to describe nonuniform  complex fluids is built based on the ideas
of DFT for simple fluids~\cite{lutsko}.
Perhaps, the first formulation of the DFT for
polyatomic fluids was given by Chandler et al.~\cite{chandler1,chandler2}.
They extended the DFT of
nonuniform simple fluids to the systems composed of polyatomic species,
considered in the framework of the RISM approach~\cite{chandler3,chandler4}.
By using Legendre transforms, these authors proved
the existence of a free energy  functional with the local densities
referring  to the  interaction sites rather than to molecular coordinates.
The methodology proposed by Chandler et al.~\cite{chandler1,chandler2}
retains mathematical simplicity of the
traditional theory for atomic fluids, but
it requires  the ``site-site'' direct correlation function as an input.

Another class of  DFT approaches for polymeric fluids
was proposed by Woodward~\cite{woodward0},
who used the concept of a weighted density approximation,
common in the DFTs for simple fluids~\cite{evans}.
Woodward applied the theory to a fluid  of particles  made of
tangentially bonded hard spheres
and borrowed the construction of weighted densities from   DFT
for hard spheres. He obtained the expression for a nonlocal free energy functional
that depends on the segment density only.
This approximation was also used to investigate
a mixture of polymers and hard spheres in slit-like pores.
The  approach by Woodward was extended to polydisperse polymeric
systems~\cite{woodward1} with the Schulz-Flory mass distribution, and to
infinitely long polymers~\cite{woodward2}. At present it is clear that
that the weighing procedure of this method is crude and requires improvements.
However, a weakness of the theory is due to the presence of an adjustable parameter
to account for  complex correlation effects.

 An important group of DF approaches has been constructed
by using the thermodynamic perturbation theory (TPT) by
Wertheim~\cite{wertheim1,wertheim2,wertheim3}.
The earliest theory of this type was proposed by
Kierlik and Rosinberg~\cite{kierlik1,kierlik2,kierlik3,phan}.
This approach  applies the intramolecular distribution function
for pairs of bonded atoms to evaluate the functional of the free
energy of the system and a more sophisticated procedure for
obtaining the weighted densities.

The theory by Kierlik and Rosinberg was refined by Yu and Wu~\cite{yuwu},
who introduced a new concept for the treatment of
intramolecular degrees of freedom.  The novelty is in the
use of the fundamental measure theory to accurately evaluate
the free energy of a hard sphere reference system and  pair cavity
distribution function at contact necessary in the TPT formulation.
The approach developed by Yu and Wu was used in several
theoretical studies of inhomogeneous polymeric systems
including surface phase transitions~\cite{brykp1}, phase equilibria
in confined geometry~\cite{brykp2,li_z1,li_z2}
and effective interactions~\cite{li_z3}.
The Yu and Wu functional  permitted  to derive
an analytical expression for the polymer-wall interfacial
tension~\cite{macdowell}, to study
star polymers~\cite{mali_sasha,xu_x1,sommer}, rod-polymer
mixtures~\cite{brykp3}, and hyperbranched polymers~\cite{cao1}.
The approach by Yu and Wu has been widely used to study
the systems involving tethered chains \cite{cao2,jiang_t,xu_x2,
borowko1,pizio1,patrykiejew,pizio2,borowko2,borowko3,borowko4,pizio3,borowko5,pizio4}.

Making a short overview of the DF approaches used
in the description of nonuniform fluids involving chain
molecules, one should mention a quite accurate TPT-based DFT approach
by Chapman et al.~\cite{chapman1,chapman2,chapman3}. It concerns
the associating fluids that can form different complexes and chains, in particular.
The DF formulation for associating monomers is supplemented by
appropriate mass action laws. The limit of complete association
is imposed at the final stage of the procedure, in contrast to the
method by Yu and Wu. On the other hand, associative approach is
useful to describe the bonding of chain molecules with a solid surface.

It should be noted that there exists a link
between DF methods and the SCFT
for systems of polyatomic molecules. Several
DFTs can be recast to the SCFT form \cite{freed0,tangfreed} with
a different degree of difficulty. Generally,
the so-called DFT-SCFT approaches rely on a
combination of computer simulations of a single chain
in an effective position dependent field that results
from nonlocal DFT~\cite{yethiraj1,chandra,muller1,muller2,muller3,muller4}.
The advantage of the DFT-SCFT approaches is that
practically all global and local properties of polymers at interfaces,
such as the center-of-mass profile or the end-to-end vector profile,
can be  calculated and analyzed~\cite{yethiraj2,cao3,chen,brykp4}.
In particular, those methods can be used to take into account
the shrinking or the swelling polymers at the surface. So far, such
calculations have been carried out only for bulk systems using
PRISM integral equation approaches~\cite{melenkevitz,grayce1,grayce2}.

Quite recently, Bryk and MacDowell proposed a new
hybrid self-consistent field DF approach
to describe the solvation effects for polymers at interfaces~\cite{brykp5}.
Using DFT of Kierlik and Rosinberg~\cite{kierlik1,kierlik2,kierlik3,phan},
they showed that when the second-order terms of Wertheim
thermodynamic perturbation approach (TPT2)~\cite{wertheim1,wertheim2,wertheim3}
are taken into account,
 the DFT-TPT2 is capable of capturing the effects of solvation of tethered chains.
Theoretical predictions were tested with respect to
Monte Carlo simulations for moderate chain lengths~\cite{brykp5}.
The results for the end-to-end distances of chain molecules
in the solvent  in the bulk phase and at the solid surface
were in a reasonable agreement with simulations.

\section{Density functional theory}\label{sec3}

As we have already mentioned in the introductory
section, different formulations of DFT are available in the
theory of inhomogeneous complex and polymeric fluids.
In this text, however, we restrict ourselves to solely one, frequently used version,
for illustrative purposes.
It contains elements of other approaches and
in the course of presentation we will address
the issues related to other DFTs.

Let us consider the fluid (indexed as 1)
in contact with the surface covered
with the end-grafted chains, $P$,~\cite{cao2,jiang_t,xu_x2,borowko1,pizio1,
patrykiejew,pizio2,borowko2,borowko3,borowko4}
(an extension to
fluid mixtures  is straightforward).
The chain consists of $M$ tangentially bonded segments.
Although
other models
involving different segments can be considered,
we restrict our attention to the case where all the segments are identical.

The connectivity of segments within a chain is provided by the
bonding potential, $V_\textrm{B}$,  defined as follows:
\begin{equation} \label{eq:1}
\exp [-\beta V_\textrm{B}({\bf R})]=\prod_{i=1}^{{M}-1}\delta \left(|{\bf r}_{i+1}-%
{\bf r}_{i}|-\sigma^{(P)}\right)\Big/4\pi \left(\sigma^{(P)}\right)^{2},
\end{equation}
where ${\bf R}\equiv (\mathbf{r}_{1},\mathbf{r}_{2},\cdots ,\mathbf{r}_{{M}})$
is the vector specifying the positions of segments and
$\sigma^{(P)}$ is the diameter of the segments.
It is assumed that all the segments and fluid molecules interact
with each other  and with solvent molecules by means of (12--6) Lennard-Jones potential,
\begin{equation} \label{eq:2}
u^{(kl)}(r)=
4 \varepsilon^{(kl)}\left[\left(\sigma^{(kl)}/r\right)^{12}-\left(\sigma^{(kl)}/r\right)^6\right],
\end{equation}
where $ \varepsilon^{(kl)}$ is the strength of interactions between
the species $k$ and $l$ and $\sigma ^{(kl)}=\left(\sigma^{(k)} +\sigma^{(l)}\right)/2$, ($k,l=P,1$).

The chains are  pinned
to the surface.
The interaction of their first segments with the surface is given by
\begin{equation}
\exp\left[-\beta v_{1}^{(P)}(z)\right]=C \delta \left(z-\sigma^{(P)}/2\right),
\label{eq:bin1}
\end{equation}
where $z$ is the distance
from the surface and $C$ is a constant.
The remaining segments of the chains, $j=2,3,\dots,M$,
as well as  the fluid molecules   interact with the surface via
a hard-wall potential,
\begin{equation}
 v^{(k)}_j(z)=\left\{
\begin{array}{ll}
 \infty, & z < \sigma^{(k)}/2, \\
0,  &  \text{otherwise},
\end{array}
\right.
\label{eq:hw}
\end{equation}
or, depending on the model used,
via the Lennard-Jones (9--3) potential,
\begin{equation}
v^{(k)}_j(z)= \varepsilon_{\mathrm{s}}^{(k)}\left[\left(z_0^{(k)}/z\right)^{9}-\left(z_0^{(k)}/z\right)^3\right].
\label{eq:lj93}
\end{equation}
In the above equation, $\varepsilon_{\mathrm{s}}^{(k)}$  is the strength
of interaction and $z_0=\sigma^{(k)}/2$.
The geometry of the system can be changed.
One can study a slit-like pore,
then the potentials (\ref{eq:bin1}), (\ref{eq:hw}) and
(\ref{eq:lj93}) must be modified, cf.~\cite{borowko1,pizio1}.

In order to present
the theory, let us introduce
the notation $\rho ^{(P)}(\mathbf{R})$ and $\rho ^{(1)}(\mathbf{r})$ for the
density profiles  of chains and fluid, respectively, and define the segment densities
$\rho _{\textrm{s}j}^{(P)}$ and the total segment density of chains, $\rho
_\textrm{s}^{(P)} $, via the relation,
\begin{equation}
\rho _\textrm{s}^{(P)}(\mathbf{r})=\sum_{j=1}^{M}\rho _{\textrm{s}j}^{(P)}(\mathbf{r}
)=\sum_{j=1}^{M}\int \rd \mathbf{R}\delta (\mathbf{r}-\mathbf{r}_{j})\rho
^{(P)}(\mathbf{R})\,.  \label{eq:6}
\end{equation}
To simplify the equations, we use the notation $\rho _\textrm{s}^{(1)}(
\mathbf{r})\equiv \rho _{\textrm{s}1}^{(1)}(\mathbf{r})\equiv \rho ^{(1)}(\mathbf{r})$.

The
system  is considered in the grand canonical ensemble.
The thermodynamic potential, $\Omega$, is as follows:
\begin{equation}
\Omega=F\left[\rho ^{(P)}(\mathbf{R}),\rho ^{(1)}(\mathbf{r})\right]+\int \rd
\mathbf{R}\rho ^{(1)}\left(\mathbf{r}\right)\left[v_1^{(1)}(\mathbf{r})-\mu \right]
+\sum_{j=1}^M\int \rho^{(P)}_{sj}(\mathbf{r})v_j^{(P)}(z),
\end{equation}%
where $F\left[\rho ^{(P)}(\mathbf{R}),\rho ^{(1)}(\mathbf{r})\right]$ is the Helmholtz
free energy functional and $\mu$ is the chemical
potential of \linebreak species~1. The functional $\Omega$ is evaluated under the constraint
\begin{equation}
\frac{1}{A}\int \rd\mathbf{r}\rho _{\textrm{s}j}^{(P)}(\mathbf{r}) =\rho_\textrm{C}\,,
 \label{eq:con}
\end{equation}
where $A$ is the surface area and $\rho_\textrm{C}$ is the
surface density of grafted chains.

The free energy functional is divided into ideal,
 $F_\textrm{id}$, and the excess, $F_\textrm{ex}$, terms.
 The excess terms are due to the hard-sphere interactions
($\textrm{hs}$), due to the connectivity of segments of chain molecules ($\textrm{c}$)
and due to attractive interactions between all the species ($\textrm{att}$),
$F=F_\textrm{id}+F_\textrm{ex}$, $F_\textrm{ex}=F_\textrm{hs}+F_\textrm{c}+F_\textrm{att}$. The configurational ideal
part of the free energy functional is as follows:
\marginpar{\em\color{red}missing \\
brackets !!!}
\begin{eqnarray}
\beta F_\textrm{id}&=&\beta \int \rd\mathbf{R}\rho ^{(P)}(\mathbf{R})V_\textrm{B}(\mathbf{R
})+\int \rd\mathbf{R}\rho ^{(P)}(\mathbf{R})\Big[\ln \left(\rho ^{(P)}(\mathbf{R}
)\right)-1\Big]\nonumber\\
&&+\int \rd\mathbf{r}\rho ^{(1)}(\mathbf{r})\Big[\ln \left(\rho ^{(1)}(\mathbf{r}
)\right)-1\Big]\,.
\end{eqnarray}
The excess free energy,  due to hard-sphere interactions,
$F_\textrm{hs}=\int \Phi _\textrm{hs}(\mathbf{r})d\mathbf{r}$,
results from
the  ``white bear'' version of the fundamental measure theory. Then, we have~\cite{yuwu}
\begin{equation}
\Phi _\textrm{hs}=-n_{0}\ln (1-n_{3})+\frac{n_{1}n_{2}-\mathbf{n}_{1}\cdot \mathbf{n
}_{2}}{1-n_{3}}+n_{2}^{3}(1-\xi ^{2})^{3}\frac{n_{3}+(1-n_{3})^{2}\ln
(1-n_{3})}{36\pi n_{3}^{2}(1-n_{3})^{2}}\;.  \label{eq:7}
\end{equation}
In the above, $\xi (\mathbf{r})=|\mathbf{n}_{2}(\mathbf{r})|/n_{2}(\mathbf{r}
) $. The definitions of
scalar, $n_{\alpha }$, for $\alpha =0,1,2,3$ and vector, $\mathbf{n}
_{\alpha }$, for $\alpha =1,2$ weighted densities are given in reference~\cite{yuwu}.

The contribution $F_\textrm{c}=\int \Phi _\textrm{c}(\mathbf{r})\rd\mathbf{r}$ is evaluated
from Wertheim's first-order TPT~\cite{wertheim1,wertheim2,wertheim3}
\begin{equation}
\Phi _{(\textrm{c})}=\frac{1-M}{M}n_{0}^{(P)}\zeta ^{(P)}\ln \left[y_{(\textrm{hs})}^{(P)}\left(\sigma
^{(P)}\right)\right]\,,  \label{eq:8}
\end{equation}
where $\zeta ^{(P)}=1-\mathbf{n}_{2}^{(P)}\cdot \mathbf{n}
_{2}^{(P)}\big/\big(n_{2}^{(P)}\big)^{2}$ and the contact value
of the cavity
function of hard spheres,
$y_{(\textrm{hs})}^{(P)}$, results from Boublik-Mansoori-Carnahan-Starling-Leland
equation of state,
\begin{equation}
y_\textrm{hs}^{(P)}(\sigma ^{(P)})=\frac{1}{1-n_{3}}+\frac{n_{2}\sigma ^{(P)}\zeta
}{4(1-n_{3})^{2}}+\frac{\left(n_{2}\sigma ^{(P)}\right)^{2}\zeta }{72(1-n_{3})^{3}}\;,
\label{eq:9}
\end{equation}
with $\zeta =1-\mathbf{n}_{2}\cdot \mathbf{n}_{2}/(n_{2}^{{}})^{2}$.

Finally, the attractive contribution is given by the mean-field
approximation
\begin{equation}
F_\textrm{att}=\frac{1}{2}\sum_{k,l=1,P}\int \rd\mathbf{r}\rd\mathbf{r^{\prime }}\rho
_\textrm{s}^{(k)}(\mathbf{r})\rho _\textrm{s}^{(l)}(\mathbf{r^{\prime }})u_\textrm{att}^{(kl)}(|
\mathbf{r}-\mathbf{r^{\prime }}|),
\end{equation}
where $u_\textrm{att}^{(kl)}(|\mathbf{r}-\mathbf{r^{\prime }}|)$ is defined according to
Weeks-Chandler-Andersen scheme,
\begin{equation}
u_\textrm{att}^{(kl)}(r)=\left\{
\begin{array}{ll}
-\varepsilon ^{(kl)}, & r\leqslant  2^{1/6}\sigma ^{(kl)}, \\
u^{(kl)}(r), & r>2^{1/6}\sigma ^{(kl)},
\end{array}
\right.
\end{equation}
and the diameter
is  assumed $d^{(i)}=\sigma^{(i)}$, ($i=P,1$).

The equilibrium density profiles are obtained
by minimizing $\Omega$ under the constraint (\ref{eq:con}),
\begin{equation}
 \frac{\delta \Omega}{\delta \rho^{(P)}\mathbf{R})}=0, \qquad
  \frac{\delta \Omega} {\delta \rho^{(1)}(\mathbf{r})}=0.
\end{equation}
The final equations for the segment density profile and for the fluid
density profile can be found in references~\cite{yuwu,borowko1,pizio1}.

The bulk fluid in equilibrium with the system of interest is
the fluid of Lennard-Jones particles of type~$1$. The configurational chemical
potential $\mu$ in terms of the bulk density $\rho _\textrm{b}$ of species $1$ is
\begin{equation}
\beta \mu =\ln \rho _\textrm{b}+\beta \mu _{(\textrm{hs})}+\beta \rho _\textrm{b}\int
u_\textrm{att}^{(11)}(r)\rd\mathbf{r},
\end{equation}%
where $\mu _{(\textrm{hs})}$ is the excess chemical potential of hard spheres.

The knowledge of the density profiles is crucial for the evaluation
of thermodynamic properties.
The density profiles provide the excess adsorption isotherms of
fluid species, $\Gamma_\textrm{ex}$,
\begin{equation}
\Gamma_\textrm{ex} = \int \rd z \left[\rho^{(1)}(z)-\rho_\textrm{b}\right],
\end{equation}
where the limits of integration are determined by the geometry of the
system.
The knowledge of the density profiles yields the value of $\Omega$ and thus
permits for the evaluation of the solvation
force
\begin{equation}
\beta f/(2A)= -\frac{1}{A} \frac{\partial \beta \Omega}{\partial H}=\int \rd z
\rho^{(1)}(z)\frac{\partial v_1^{(1)}(z)}{\partial z}
+\sum_{j=1}^M\int \rd z
\rho^{(P)}_{\textrm{s}j}(z)\frac{\partial v^{(P)}_j(z)}{\partial z}\,.
\end{equation}
The solvation force between the surfaces is
a measurable property. It comes out from
the experiments performed
by using a surface force apparatus, see e.g.,~\cite{israelichvili}.

Our final comment concerns another important property. Upon fluid adsorption,
the average height of tethered chains changes in a nontrivial way. In order to
describe this property, we use the normalized
first moment,
\begin{equation}
h=a\int \rd z z \rho^{(P)}_\textrm{s}(z)\bigg/\int \rd z \rho^{(P)}_\textrm{s}(z).
\end{equation}
Depending on  the physical argument, the coefficient $a$ is taken
equal to $8/3$ (cf.~\cite{kreer,dimitrov}) or $a = 2$~\cite{pastorino}.

The theory presented above  permits for several extensions.
In particular, it is possible
to consider either tethered
 copolymers \cite{sommer1} or
 chain layers of different species.
 Flexibility of bonding between segments can
be modified as well, assuming, for example,
preferential angles
for bonding between certain groups of segments~\cite{jiang_j1}.
 It is also possible to consider
the branched polymers and to change the identity of segments (and their sequence).

\section{Applications of the density functional approaches}\label{sec4}

In this section,
we would like to focus on a few  results
of DFTs.
In general, one needs to be convinced that
the obtained density profiles and the values of $\Omega$
are accurate for
different values of parameters describing the system.
The profiles yield the
excess adsorption
and the values of $\Omega$ allow for the investigation of
the phase behavior.

We proceed  to a  discussion of
density profiles.
The configuration of end-grafted polymers in a good
solvent has been well documented. Specifically, the SCF theory by Milner et al.~\cite{milner}
based on the ideas of Alexander~\cite{alexander}
and de Gennes~\cite{gennes}, lead to  conclusions
that at a moderate surface grafting density, the chains exhibit
parabolic  density profiles. This result has been supported
by simulation data.
We focus on exploring the possibilities of DFTs and the
DFT by Chapman et al.~\cite{chapman3},
denominated as iSAFT (interfacial statistical associating fluid theory).
As we have already mentioned, this approach  slightly differs from the theory  described above 
by the treatment of
bonding of chain molecules to the solid surface.

\begin{figure}[h]
\begin{center}
\includegraphics[width=0.46\textwidth]{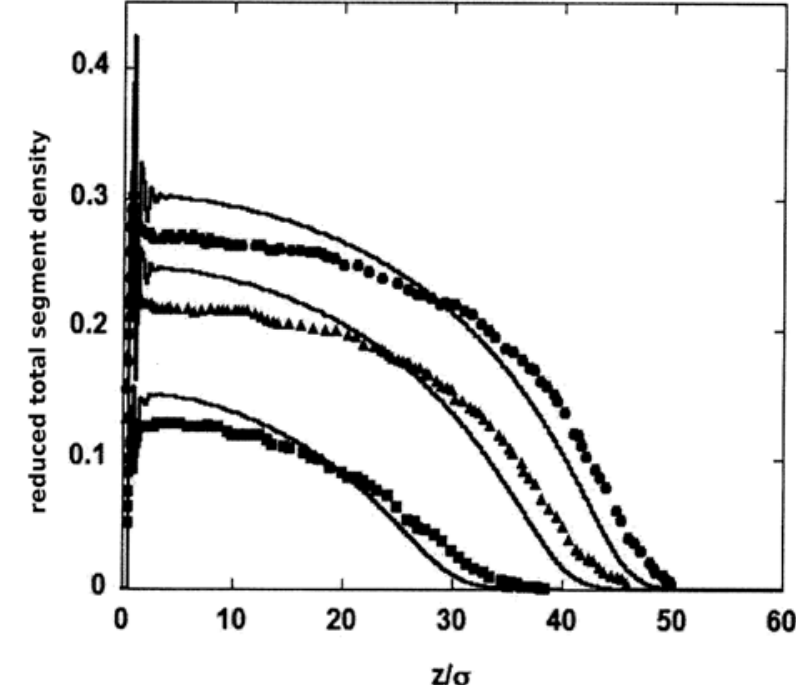}
\hspace{5mm}
\includegraphics[width=0.45\textwidth]{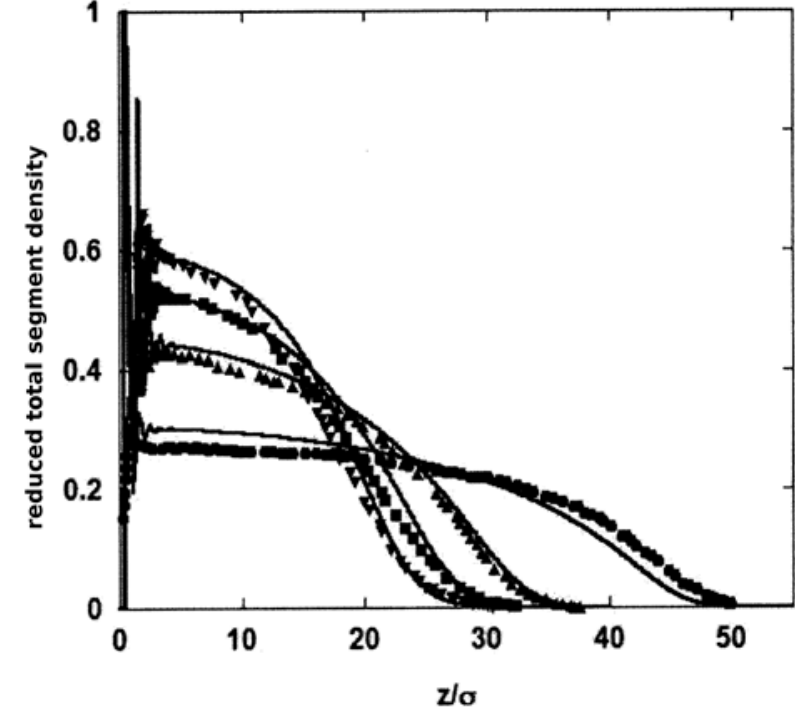}
\end{center}
\caption{Left hand panel: segment density profiles of chains of $M=100$ segments tethered to a hard wall
at a grafting density $\rho_\textrm{C} \sigma^2$=0.03 (squares), 0.07 (triangles),
and 0.1 (circles).
Right hand panel: segment density profiles for
a grafting
density $\rho_\textrm{C} \sigma^2=0.1$ in the presence of implicit solvent
(circles) or explicit polymer solvent with $M_f=2$ (triangles up),
$M_f=5$ (squares), or $M_f=10$ (triangles down).
Symbols are the simulation
results~\cite{grest}, and the curves result from iSAFT.
The total segment density of  fluid is
$\rho_\textbf{b}\sigma^3=0.682$.
Reprinted with permission from reference~\cite{chapman3}. Copyright AIP.
}
\label{fig1}
\protect
\end{figure}

The first illustration~\cite{chapman3} shows the structure of
athermal chains with $M=100$ hard-sphere segments  tethered to a  hard wall.
The density profile is parabolic (left hand panel of figure~\ref{fig1}).
At a high grafting density, the deviations from a parabolic shape are not big.
In general, this DFT overestimates the density profile values close
to the wall and underestimates the profile further from the wall.
However, the difference between simulation data~\cite{grest} and theory is not large.
In addition,  Chapman et al.~\cite{chapman3} showed that
the DFT yields predictions similar to the SCF approach~\cite{zhulina1}.
The right hand panel of figure~\ref{fig1} shows the density profiles
of chains tethered to a planar surface in the presence of free polymer solution
of short chains resulting from  iSAFT theory and compares them  with simulation
results and with the profiles for an implicit solvent.
The density profiles for the models with explicit solvent significantly differ
from the implicit solvent case. The segments of solvent chains interact via
a purely repulsive potential. Consequently, the observed  effect  is entropic.
In an explicit solvent, the solvent molecules compress the chains and
the layer height is reduced compared to an implicit solvent $N_\textrm{c}$ kind of linear
homopolymers
case.
The total solvent density is high, see the figure caption.
The compression  increases with an increase
of the number of
segments of the solvent, showing that  penetration of longer ($M=10$) free
chains into the tethered layer decreases, compared to shorter chains.
Thus, the chain solvent effect is to reduce the tethered layer height 
and to
make it denser in the vicinity of the solid surface.
The observations from the DFT
are in agreement with  computer simulations~\cite{grest}.

\subsection{Phase transitions}

After evaluating the density profiles, one can try to
describe the thermodynamics of adsorption
and to analyze  the phase behavior.
Such problems were studied by Bor{\'o}wko et al.~\cite{borowko3}, who
considered a monoatomic fluid
in contact with modified and non-modified solid surfaces. The fluid-wall interaction
was given by the Lennard-Jones (9,3) potential.
The surface was modified by short (5 segments) chain molecules.
The model assumed the same interactions $\varepsilon=\varepsilon^{(kl)}$  between all
species, the diameters of segments and of fluid particles were the same and
$\varepsilon_\textrm{s}/\varepsilon=8$; $\varepsilon_\textrm{s}=\varepsilon_\textrm{s}^{(P)}=\varepsilon_\textrm{s}^{(1)}$.
The
DF approach described in section~\ref{sec3} was used.
\begin{figure}[h]
\begin{center}
\includegraphics[width=11.5cm,clip]{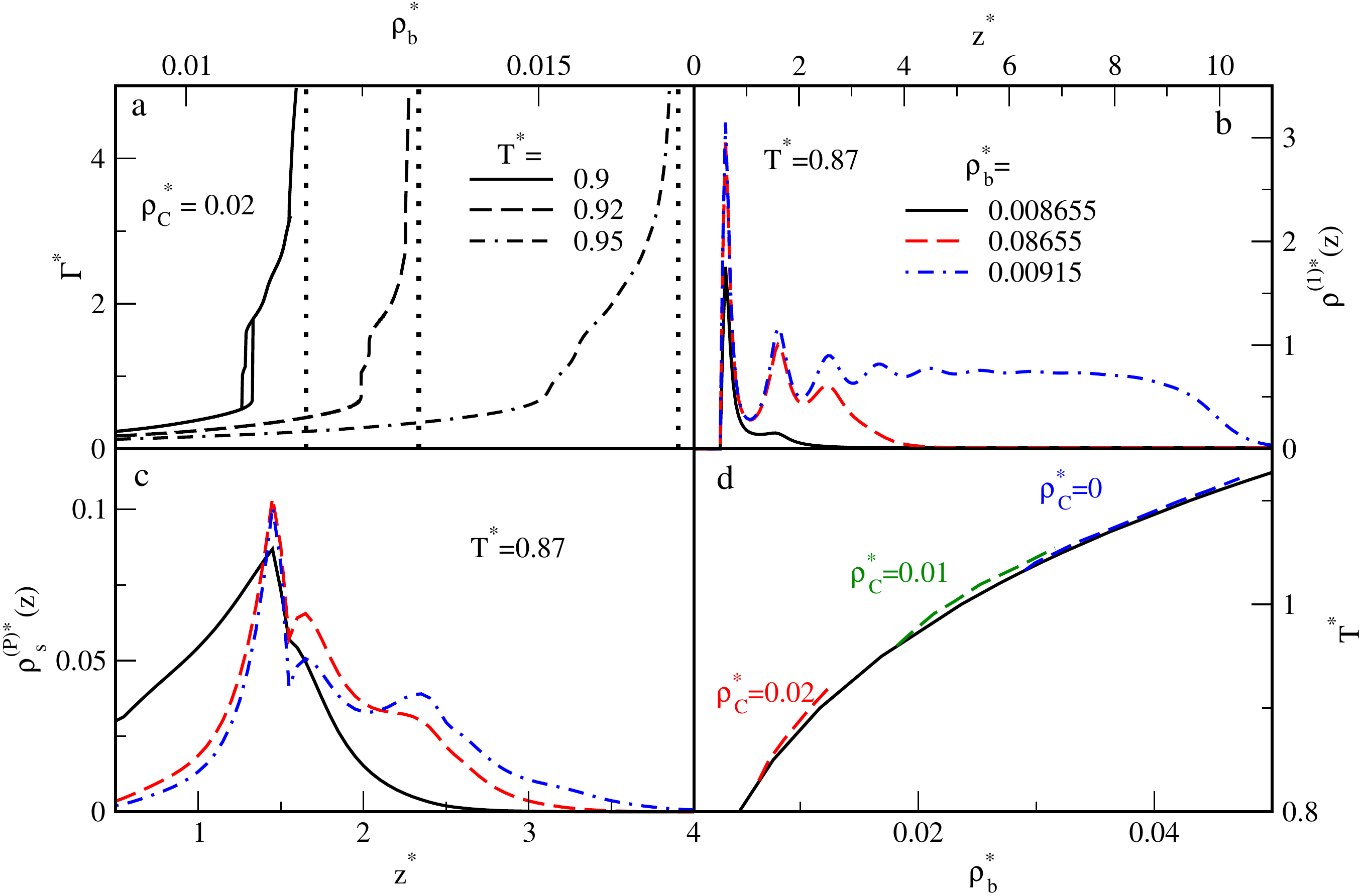}
\end{center}
\caption{(Color online) (a): Adsorption isotherms on a modified surface with the grafting
density $\rho^*_\textrm{C}=0.02$ at different temperatures. Dotted vertical lines denote the
bulk liquid-vapor transition. Dashed vertical lines
denote the equilibrium thin-thick film transition.
(b) and (c): Fluid (b) and segment (c) local densities at $T^*=0.87$ and
for bulk fluid densities given in the (b). (d): The surface
phase diagram. Solid line is the bulk phase diagram, dashed lines
are the wetting diagrams for $\rho_\textrm{C}^*=0.01$, 0.2 and for non-modified
surface,  $\rho_\textrm{C}^*=0$.
}
\label{fig2}
\protect
\end{figure}

A set of adsorption isotherms,
at different reduced temperatures, $T^*=kT/\varepsilon$,
is shown in figure~\ref{fig2}~(a). At high temperatures,
the adsorption isotherms, $\Gamma^*=\Gamma\sigma^2$, monotonously
increase with bulk
density, $\rho_\textrm{b}^*=\rho_\textrm{b}\sigma^3$.
However, at lower temperatures, they
exhibit  hysteresis loops.
The first-order phase transition point
at each temperature was found from the crossing of two
branches of the thermodynamic
potential, $\Omega$, as functions of
chemical potential of the fluid.
Dashed lines   in  figure~\ref{fig2}~(a) show the bulk densities
at the transition points. The changes in
the microscopic structure  upon transition
can be seen on the density profiles, see figure~\ref{fig2}~(b)
for $T^*=0.87$. One observes a transition from a thin to
a thick fluid film that is accompanied by changes of the
tethered layer  structure [figure~\ref{fig2}~(c)].
It is evident that
the jump of the fluid film thickness occurs within the tethered polymer
layer.
The changes of the structure of
tethered chains are shown in [figure~\ref{fig2}~(c)]. The growth of the fluid
film causes the swelling of tethered  chains.

In order to characterize these phenomena  it is necessary to construct the phase diagram.
Numerical calculations  revealed
that for the system  without
tethered chains, the wetting temperature
is $T^*_\textrm{w}=kT_\textrm{w}/\varepsilon= 1.03$, whereas the surface critical temperature
is  $T^*_\textrm{sc}=kT_\textrm{sc}/\varepsilon=1.125$. Upon introducing even a small amount of  chains,
one observes a shift of both characteristic temperatures
to lower values. At the grafting density $\rho_\textrm{C}^*=\rho_\textrm{c}\sigma^2=0.01$, the values of
$T^*_\textrm{w}$ and $T^*_\textrm{sc}$  are
$T^*_\textrm{w}=0.83$, and $T^*_\textrm{sc}=0.92$, respectively, see figure~\ref{fig2}~(d).

Similarly to the changes of the wetting properties, one can investigate
layering transitions. The layering transitions are usually
discussed for fluids in contact with
strongly adsorbing surfaces. The aim
is to explore how tethered chains change the scenarios for layering transitions.
In particular, whether it is possible to suppress the layering of a fluid
or to eliminate some transitions.
However, for some systems, one can also
observe  new layering  transitions as a consequence of
chemical modification of a weakly attractive surface by strongly
attractive chains.
The important factors effecting the changes of the phase behavior
are the differences in the interaction energies between the bare
surface and the segments of chains and the entropic
effects due to the changes in the structure of chains.
The parameters  such as the number of segments
in the chains and  grafting density are of importance
as well. These factors permit to tune the phase behavior
in terms of the formation or destruction of consecutive layers.
In the case of adsorption in pores, the modification
of the pore walls effects the capillary condensation. In this case,
the ratio between the chain length and the pore width becomes a new
important parameter.
The chains can also enhance or inhibit layering transitions that may
preceed the capillary condensation.
All these problems were discussed
in references~\cite{pizio1,patrykiejew}.

If the binding of the
terminal segments of chains to
the walls of a slit-like pore is not strong, a new kind of phase
transition can take place. This transition is connected with the change
of the symmetry of the distribution of chains and fluid molecules
inside the pore. Upon the adsorption of a fluid,   a part
of weakly tethered chains can be
``wiped out'' from one wall and bonded to the opposite one.
Consequently, the distribution of the chains inside the pore becomes
non-symmetric. As a result, the symmetry of the distribution
of fluid molecules with respect to the pore center is also broken.
A further increase of the fluid density may induce a reentrant behavior
manifested in the recovery of the symmetry.
 Such changes of symmetry
may occur as first- or second-order
transitions~\cite{pizio3}.

The symmetry breaking transitions are also observed for slit-like pores
modified with chains whose two ends are permanently attached to the opposing walls,
provided that the length of chains is sufficiently  larger
than the pore width~\cite{borowko5,pizio4}.
Such chains are often called pillars.
In this case, the inner segments of the chain molecules
can move towards one of the pore walls.
This problem was discussed in references~\cite{patrykiejew,borowko3}.

The phase transitions, associated with the changes of
the symmetry of local densities can be observed if the chains are much longer than
the pore width. For chains of the length comparable to the pore
width,  such transitions do not occur~\cite{borowko5}. Moreover, the symmetry
changes are possible if the surface density of the chains is moderate.
In the limiting case of a very small grafting density, the
phase transformations in modified and non-modified pores
are similar. In another limiting case of a very high grafting density,
all phase transitions within the confined fluid disappear.
The values
of the fluid-wall energy parameters are also important. In
the case of strongly attractive adsorbing potential, the symmetry
breaking is linked with the layering (that takes place at
one wall). For weaker adsorbing potentials, the growth of
a non-symmetric film at one wall resembles the growth of a
film after the prewetting jump. Finally, for a very weak adsorbing
potential, the symmetry breaking does not occur.
In the case of a strongly adsorbing surface, the first-order
symmetry change transitions end up at the tricritical points from
which the second-order transition lines depart. There exists
a triple point between the ``return-to-symmetry'' transition and
the capillary condensation envelope. In the case of weaker
adsorbing potentials, however, the second-order symmetry
change line joins the capillary condensation part of the phase
diagram at the critical end point. In such cases, a symmetry
recovery at lower temperatures takes place during capillary
condensation.

The phase behavior becomes even more complicated if the adsorption of a
binary mixture inside the modified pore is considered. If the
mixture exhibits a demixing in the bulk phase, then demixing transitions
inside the pore, or even within particular layers formed upon
the layering transitions, compete with the symmetry breaking. This competition
may lead to complex phase diagrams that include
 critical points, lower and upper critical end points,
tricritical points, $\lambda$-lines connected with demixing
as well as  $\lambda$-lines connected with symmetry breaking
and symmetry recovery transitions. Some of these problems were
discussed in reference~\cite{pizio4}.

\subsection{Solvation force}
There are several works focused on the study of the effective interaction
between surfaces coated with chains. In particular, Wu et al.~\cite{cao2}
 studied
the solvation force between two surfaces covered with  telechelic
chains. Telechelic chains are chain molecules with certain functional groups
at both ends. Thus, the end segments of a polymer can bind to a single surface
simultaneously to form a loop, or  to the opposite surfaces to form a bridge-like
structure. The backbone segments in between the two terminal groups are
chemically inert.
Experimental and theoretical studies show that the interaction
between highly stretched telechelic chains is mostly repulsive, in close
similarity to the behavior observed for the system of  two layers of singly-tethered
chains. If the two layers of  telechelic chains are put almost at
contact, i.e., if their separation distance is close to the double height of
an isolated chain, a weak attractive force is observed. The origin and the magnitude
of this attraction has been discussed by different
authors~\cite{eiser,wijmans,zhulina2}.
\begin{figure}[htb]
\begin{center}
\includegraphics[width=8.0cm]{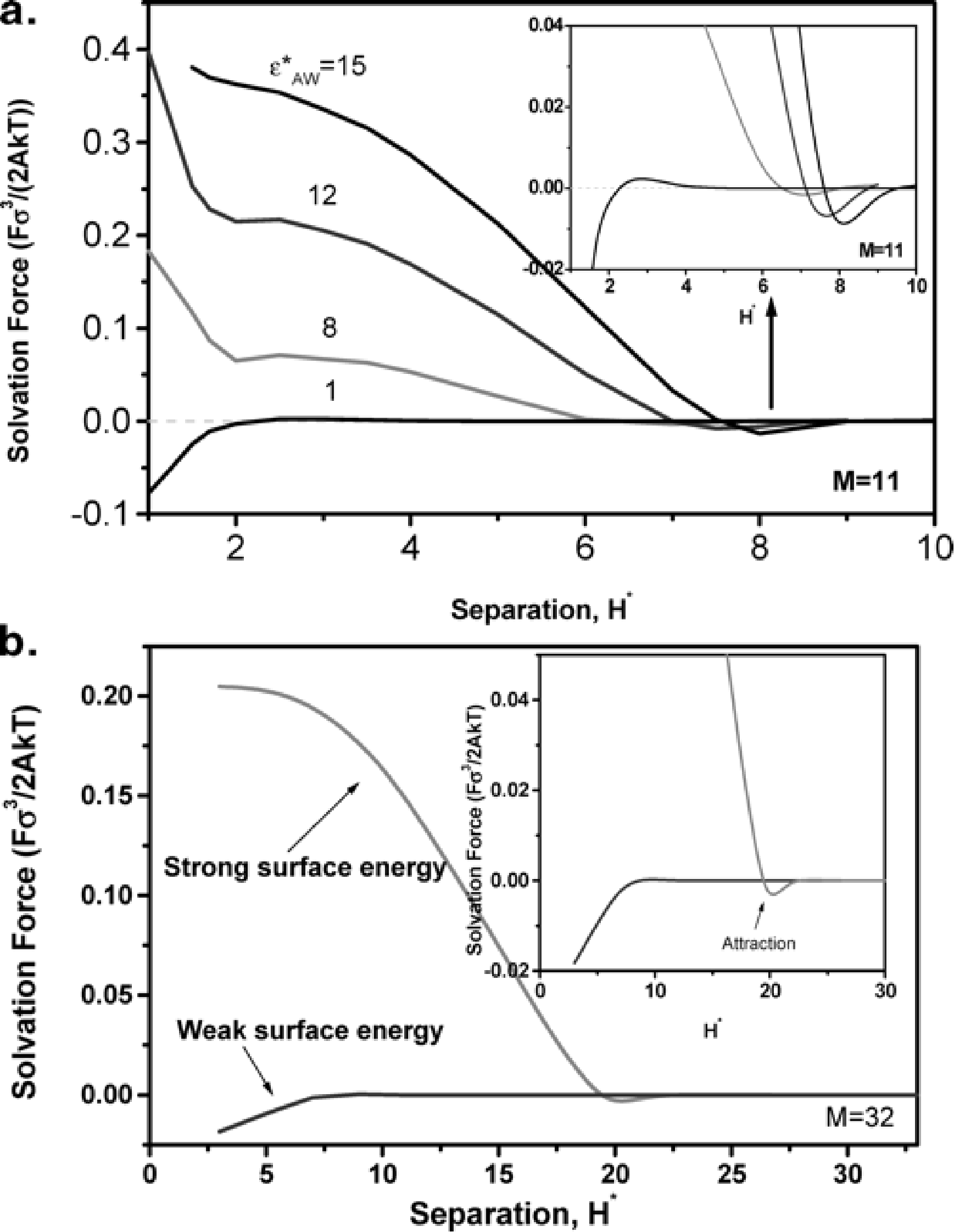}
\end{center}
\caption{Solvation force between two surfaces mediated by
telechelic polymers: (a) chain length $M=11$, and (b) $M=32$,
($M$ is the number of segments). The inset
plots are magnified views near the classical contact, i.e., when the separation between the two surfaces is
about twice the brush thickness.
Reprinted with permission from reference~\cite{cao2}. Copyright ACS.
}
\protect
\label{fig6}
\end{figure}
Similarly to the DFT by Chapman et al.~\cite{chapman1,chapman2,chapman3},
the approach by Wu et al.~\cite{cao2}
involves a parameter describing the strength of an attractive interaction
between terminal segments and the surface, surface-adhesive energy, $\varepsilon_\textrm{Aw}$.
In figure~\ref{fig6}, taken from~\cite{cao2}, the solvation force between two identical
surfaces covered by telechelic chains is shown. The
dependence of the solvation force  on the value of $\varepsilon_\textrm{Aw}$
and on the number of segments, $M$,
was studied.
The segment packing fraction in the bulk phase of chains
was rather low, $\rho_\textrm{b}^*=0.05$.
If the surface-adhesive energy increases from $\varepsilon_\textrm{Aw}/kT=1$, to
8, 12 and 15, the solvation force exhibits changes from a short-range
attraction to a long-range repulsion. At the lowest energy $\varepsilon_\textrm{Aw}/kT$, the solvation
force is similar to that for nonadsorbing polymers, $\varepsilon_\textrm{Aw}/kT=0$;
the attractive short-range part of the
force is due to the entropic depletion of chains from the surface. At
the highest value of  $\varepsilon_\textrm{Aw}/kT$, the repulsive force arises from the interaction
between tethered telechelic chains. The strength of repulsion depends
on the energy $\varepsilon_\textrm{Aw}/kT$. The magnified view of the solvation force
shown in the inset  indicates that in a nonadsorbing regime (small $\varepsilon_\textrm{Aw}/kT$),
the theory  leads to entropic depletion, as well as to a weak, secondary repulsion between surfaces. If the polymers are strongly
adsorbed, the theory predicts the existence of a weak attraction between telechelic
chains at contact. Moreover, the location of the attraction
shows that the chains are highly stretched.

DF observations are in agreement
with experimental findings and other theories.
 If the chain length
increases, similar conclusions concerning the behavior of the solvation
force preserve. A more detailed discussion can be found in reference~\cite{cao2}.
To summarize our illustrations in this part, it is worth mentioning that
similar successful developments have been used for the description of
inhomogenous fluid mixtures, tethered chains of copolymer type and
other related models with nonelectrostatic interparticle interactions.
We proceed now to another class of systems,  namely to the models with long-range
Coulomb interaction.

\section{Electrostatic density functionals}\label{sec5}

An ample set of problems of interest for laboratory and for applications
involves fluids of charged particles, as well as tethered polyelectrolytes and polyampholytes.
In the case of tethered polyelectrolyte chains, the
presence of long-range electrostatic interactions makes
the properties of such systems  much more complicated than those
corresponding to neutral grafted chains. Since the pioneering
works by Pincus \cite{pincus}, Balastre et al. \cite{balastre}
and Borisov et al. \cite{borisov},
numerous theoretical and experimental studies have been
reported concerning the structure and surface properties of
charged grafted chains.
Similarly to the case of uncharged chains, the
theoretical investigations were primarily based
on the scaling analyses and the self-consistent mean-field
theories that all rely on a number of simplifications \cite{pincus,borisov,ross,zh1,zh2,zh3}.
If the number and positions of charges
at each chain are fixed, theoretical and
experimental investigations elucidate the existence of different brush
regimes that depend on the grafting density, degree of
dissociation, and ionic strength.
However, more recent experiments by Wu et al. \cite{wuwu}
and Currie et al. \cite{currie} have shown
that the predicted scaling relations do not necessarily hold
throughout the entire range of the grafting densities and
ionic strengths.
Recently, Witte et al.~\cite{wittee} developed a new version of a SCF approach
from the grand canonical partition function.
The theory properly accounts for the local nature of the charge equilibrium
and captures the basic behaviors of grafted polyelectrolytes. In agreement with
recent experiments, the theory predicts that the scaling of grafted chains height
with grafting density can be qualitatively different
at intermediate chain lengths from that predicted by scaling arguments.
This difference was attributed
to the relative strength of electrostatic  interactions
compared to finite segment size packing constraints.
On the other hand, the trend of decreasing chains height with an increasing grafting density
predicted by the
scaling analysis is recovered for large molecular weight polymers immersed in a solution of a very weak ionic strength.

One should also mention here the SCF approach \cite{gong} aimed at predicting
the structural properties of polyelectrolytes
grafted via one of their ends to solid surfaces which
explicitly incorporates the acid-base equilibrium
responsible for the charge regulation of the polyelectrolyte groups,
as well as the conformations, size, shape, and charge
distributions of all the molecular species present.
The rest of that theory was compared with experimental
observations for the height of the layer as a function of ionic strength for
different chain molecular weight and grafting densities, and a good agreement was found.

Similarly to the case of uncharged chains,
DFT provides an alternative
method  to traditional scaling and mean-field theories by
taking into account both short- and long-range correlations in a
self-consistent manner
\cite{wuannu,liwuli}.
In order to illustrate the current  developments
of the DFT for this type of systems, we would like to briefly consider
the adsorption of an electrolyte
solution on the solid surface
modified by grafted chain molecules containing charged segments.
Commonly,
an electrolyte solution is modelled in the framework of
the primitive model, i.e., the fluid is considered as a mixture of hard-spheres
interacting via Coulomb potential. The same form of the potential
is assumed for the interaction between the fluid species and charged segments
of the tethered chains,
\begin{equation}
u_{ij}^{(\eta\alpha)}(r)=\left\{
\begin{array}{ll}
\infty , &  \qquad r<\sigma,\\[1ex]
\displaystyle\frac{{e}^{2}Z_{i}^{(\eta )}Z_{j}^{(\alpha) }}{4\pi \epsilon\epsilon_0}\frac{1}{r}, & \qquad r>\sigma,%
\end{array}%
\right.
\end{equation}
where the superscripts $\eta,\alpha = P, 1, 2$, and $Z^{\alpha}_i$ is the valency
of ions or of the charged segments. The solvent is treated as
a continuum with a given relative dielectric permittivity,
$\epsilon$ ($\epsilon_0$  is the permittivity of the vacuum).

The interaction between the ions and the wall, as well
as between the charged segments of the chains and the wall, contains
the electrostatic term,
\begin{equation}
\beta v_{j,\textrm{el}}^{(\alpha)}(z)=-2\pi l_\textrm{B} Q Z^{(\alpha)}_j  z\,,
\label{eq:2b}
\end{equation}
in addition to  non-electrostatic interaction.
Here, $Qe$ is the surface charge density on the wall
and $l_\textrm{B}= e^2/(4\pi kT \epsilon\epsilon_0)$ is the Bjerrum length.
The bulk densities of ionic species of the solution must
satisfy the electroneutrality condition
$Z_1^{(1)}\rho_\textrm{b}^{(1)}+Z_1^{(2)}\rho_\textrm{b}^{(2)}=0$.
The electroneutrality holds for
the entire system and the charge of tethered chains is
neutralized by charges of counterions.

The distribution
of all charges in the system satisfies the Poisson
equation,
\begin{equation}
 \nabla^2\Psi(z)=-\frac{1}{\epsilon\epsilon_0}q(z),
\label{eq:poi}
\end{equation}
where $\Psi(z)$ is the electrostatic potential. The solution of this
equation  requires boundary conditions.
For a fluid in contact with a single (modified) wall at $z=0$,
we can specify the value of the electrostatic potential at the wall,
setting  $V_0=\Psi(z=0)$.
The wall charge follows then from the electroneutrality condition,
\begin{equation}
 Qe+\int \rd z \ q(z) = 0,
\end{equation}
where
$
 q(z)/e=\sum_{j=1}^M Z^{(P)}_j\rho^{(P)}_{j}(z)+\sum_{\eta=1,2}Z^{(\eta)}_1\rho^{(\eta)}(z).
$
One can work either under the condition of a constant $Q$ or $V_0$.

Similarly to the previous section, the principal issue is in the
determination of the density profiles
of fluid species and of tethered chains segments by minimization of the
grand thermodynamic potential. In the present problem, the excess grand
potential is given by
\begin{equation}
 \Omega= F+\int \rd\mathbf{r} \sum_{j=1}^Mv^{(\textrm{C})}(z_j)\rho^{(\textrm{C})}_\textrm{sj}(z)+
\sum_{\eta=1,2}\int \rd\mathbf{r}  \left[v^{(\eta)}_1(z)\rho^{(\eta)}(z) -\mu_{\eta}\right]
+\int \rd \,\mathbf{r}q(z)\Psi(z).
\end{equation}
In contrast to  section~\ref{sec3} the
free energy $F$ also contains  the term due to
the coupling between electrostatic and hard sphere  interactions~\cite{wang_z1,wang_z2}.
It is given by
\begin{equation}
 F_\textrm{e}/kT =-\frac {1}{2} \sum_{\alpha,\eta, i,j}
\int \rd{\mathbf r} \rd {\mathbf r}' \Delta \bar c^{(\alpha\eta)}_{ij} (\mathbf{r},\mathbf{r}') \Delta \rho_{\alpha,i}(\mathbf{r})
\Delta \rho_{\eta,j}(\mathbf{r}').
\label{eq:fee}
\end{equation}
The summation in equation~(\ref{eq:fee}) is carried out over cations, anions and
the segments $(i,j=1,2,\dots,M)$ of the chains,
 $\Delta \rho_{P, i}(\mathbf{r})=\rho^{(P)}_{i}(\mathbf{r})$,
$\Delta \rho_{1, i}(\mathbf{r})=\rho^{(1)}(\mathbf{r})-\rho_\textrm{b}^{(1)}$ and $\Delta \rho_{2, i}(\mathbf{r})=
\rho^{(2)}(\mathbf{r})-\rho_\textrm{b}^{(2)}$.
The functions $\Delta \bar c^{(\alpha\eta)}_{ij}$ are short-range parts
of electrostatic two-particle direct correlation functions, that
 can be taken
from the analytic solution of the mean spherical approximation for a restricted
primitive model of electrolyte solutions and weighted correlation approach by
Wang et al.~\cite{wang_z1,wang_z2}.

Another difference between the systems involving
charged and uncharged segments
is that the free energy term describing the connectivity of chains is modified.
According to Wertheim's theory \cite {wertheim1,wertheim2,wertheim3} for chains built of
tangentially jointed hard spheres, the free energy contribution
$F_\textrm{c}$ (cf. section.~3)
involves the contact value of the cavity function of the
reference system of hard spheres, $y_\textrm{hs}^{(P)}$.
For uncharged systems, a key assumption of the DFT
is that  the contact value of the segment-segment cavity function  is represented by
that corresponding to a fluid composed of hard spheres, $y_\textrm{hs}^{(P)}$.  In the case
of chains built of charged segments, an approximation
for the cavity function includes the effects of screened segment-segment
electrostatic interactions and takes the form~\cite{wang_l,jiang_feng},
\begin{equation}
 y_{j,j+1}(\sigma)=
y_\textrm{hs}^{(P)}(\sigma)
\exp\left[-\frac{1}{T^*}\frac{Z_j^{(P)}Z_{j+1}^{(P)}\left(2\Gamma\sigma+\Gamma^2\sigma^2\right)}
{(1+\Gamma\sigma)^2}\right].
\end{equation}
In the above,  $\Gamma=\left(\sqrt{1+2\kappa\sigma} -1\right)/2\sigma$,
$T^*=\sigma/l_\textrm{B}$ is the ``electrostatic'' reduced temperature  and
\begin{equation}
 \kappa^2=(4\pi l_\textrm{B})\left[\sum_{j=1}^M \left(Z^{(P)}_j\right)^2 n^{(P)}_{0j}(z) +
\sum_{\eta=1,2}\left(Z^{(\eta)}_1\right)^2 n^{(\eta)}_0(z)\right]
\end{equation}
is the square of the ``local'' screening parameter~\cite{wang_z1,wang_z2}.
The application of the above approximation for chains  with different
charge sequences under different thermodynamic conditions
leads to qualitatively correct results~\cite{jiang_feng}.

\section{Selected applications of the electrostatic density fun\-ctio\-nal app\-roach}\label{sec6}

In general, tethered polyelectrolytes respond to external
stimuli such as the properties of electrolyte solutions,
density, ionic strength, temperature,
see, references~\cite{sach1,sach2,sach3,sach4},
as well  external electric field.
There exist a vast literature dedicated to experimental, simulation
and theoretical studies of such responsive chains.
Among various external stimuli,
electric field is the one of much interest because it can induce
 changes of the tethered layer height~\cite{sach2,sach5,sach6}
and consequently yields changes of, e.g., permeability, transport properties and
electric conductance of
nanochannels~\cite{sach7}.

\begin{figure}[htb]
\begin{center}
\includegraphics[height=5.5cm,clip]{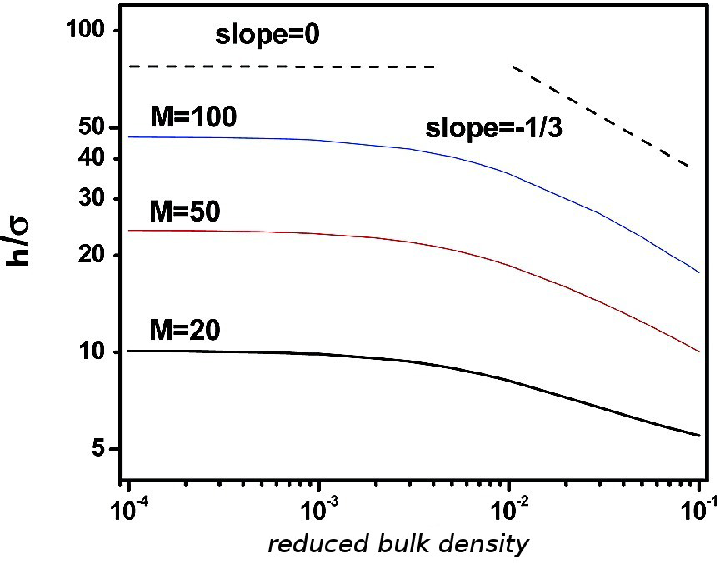}
\hspace{5mm}
\includegraphics[height=5.5cm,clip]{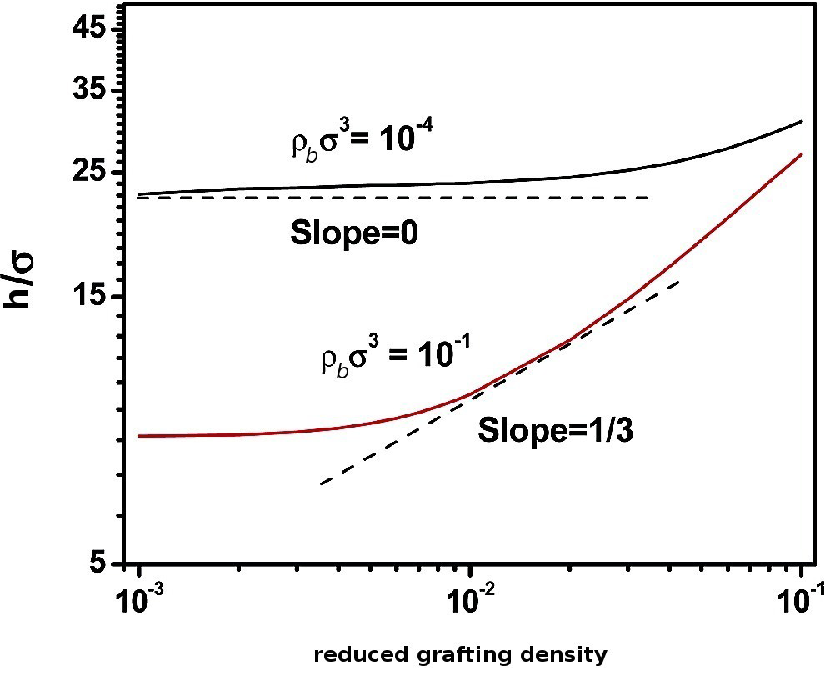}
\end{center}
\caption{(Color online) Left hand panel. Dependence of the tethered layer thickness on salt concentration for chain length
$M=20, \, 50, \, 100$. The reduced grafting density is $\rho_\textrm{C}\sigma^2=0.01$. Dashed lines
are the predictions of the scaling laws.
Right hand panel. Dependence of the tethered layer thickness on the grafting density in the
osmotic regime ($\rho_\textrm{b}\sigma^3=10^{-4}$) and in the salted-brush
regime ($\rho_\textrm{b}\sigma^3=0.1$). The chain length is $M=50$, dashed lines
are the predictions of the scaling laws.
Reprinted with permission from reference~\cite{jiang_t}. Copyright ACS.
}
\protect
\label{fig7}
\end{figure}

Upon putting  an ionic fluid in contact with the modified surface,
the height of the tethered polyelectrolyte changes
depending on different factors.
A set of results concerning these trends from the
DF approach by Jiang et al. \cite{jiang_t} is
given in figure~\ref{fig7}.
In this particular case, the polyelectrolyte grafted at the uncharged solid surface
consists of univalent negatively charged spherical segments,
whereas
the fluid (salt) is a mixture of cations and anions of the same
valency. For simplicity, the diameters of ions and
segments are assumed to be equal. At a given chain length, the height
of polyelectrolyte layer practically does not depend on the salt density in the
interval of low salt densities. This is the so-called osmotic-brush regime and
this type of behavior was first established theoretically, see e.g.,~\cite{pincus}
and confirmed experimentally~\cite{balastre}.
By contrast, at higher salt concentrations, the results shown in the figure
are in accordance with the salted-brush regime, in which the brush thickness
obeys the power law, established previously
in references~\cite{pincus,borisov}.
In the unifying picture for a wide range of salt concentrations,
one can say that polyelectrolyte chains undergo a smooth
transition from ``osmotic sticks'' to salted coils. The DFT
reproduces the previously established scaling laws for long chains. Moreover,
for shorter chains, e.g., for $M=20$, the DFT provides very good estimates for the
tethered layer height on salt density in close similarity to other approaches.

Another result concerns the dependence of the height of chains on the
grafting density. Although the grafting density
in the previous example was low, this value still yields
the brush rather than mushroom-type structure,
as it can be deduced from the values in the figure. The grafting density
parameter effects the height of the chains in a nontrivial manner. Namely, in the
osmotic regime, the height of the chains increases slightly with an increase
of grafting density showing deviation from the scaling law behavior.
Such a growth is in accordance with computer simulations and recent
experimental findings~\cite{romet,ahrens}.

This ``nonlinear osmotic'' regime (i.e., when
the concentration of counterions inside the tethered
chains is lower than in the bulk fluid) is well described by DFT.  Moreover,
the slope of the curve changes with
the grafting density, or, in other words, there is no simple scaling relation
between two variables of the figure at low salt concentration.
In the  ``salted-brush'' regime, the height of the chains grows slowly until the
separation between the grafting points of polyions becomes small,
such that at a higher grafting density, the height of the chains growth obeys a
quasi-power law. In fact, the emerging picture is richer and gives
the details that are not intrinsically provided by simple scaling laws.

Previous examples have concerned the polyelectrolytes tethered
at uncharged surfaces. However, there is an increasing interest
in developing nanofluidic systems that can control various properties,
including transport of ions and molecules in aqueous solutions
in charged nanochannels.
The next example shows
the effects of non-zero electrostatic potential at the surface
on the distribution of chain segments and of ions
in the interfacial region. One would expect that the electric double layer
will
be substantially affected by the presence of tethered chains.
Our presentation concerns
the primitive model of electrolyte. Moreover,
we assume that tethered chains consist of uncharged
segments of the same diameter as fluid ions. Two walls covered
with tethered chains constitute  a slit-like pore. We apply the theory
briefly described in the first part of this section.
The walls are  charged, and the charge is determined
by the electrostatic potential on the wall $V^*=eV_0/kT$.

\begin{figure}[htb]
\begin{center}
\includegraphics[height=6.5cm,clip]{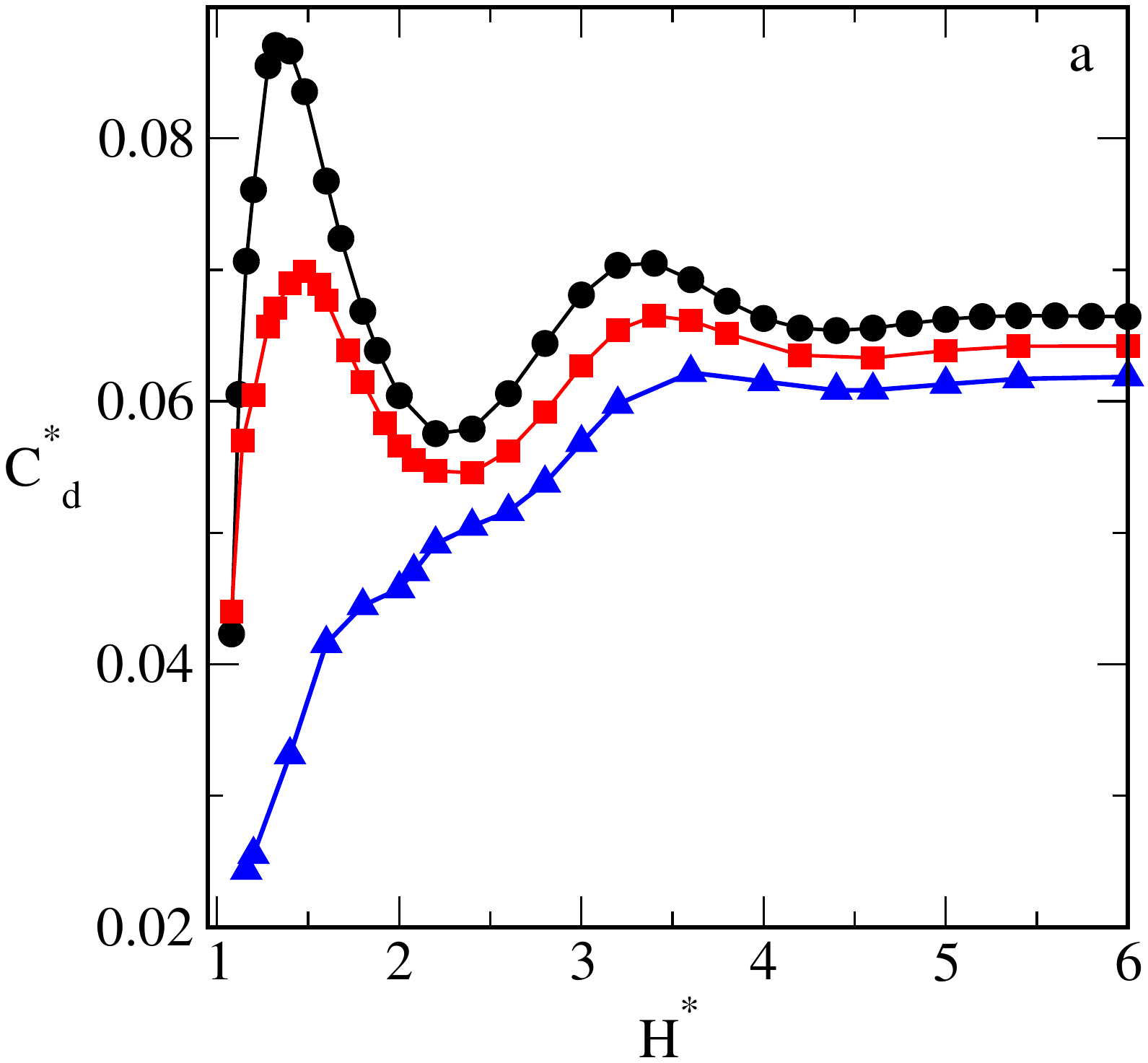}
\hspace{2mm}
\includegraphics[height=6.5cm,clip]{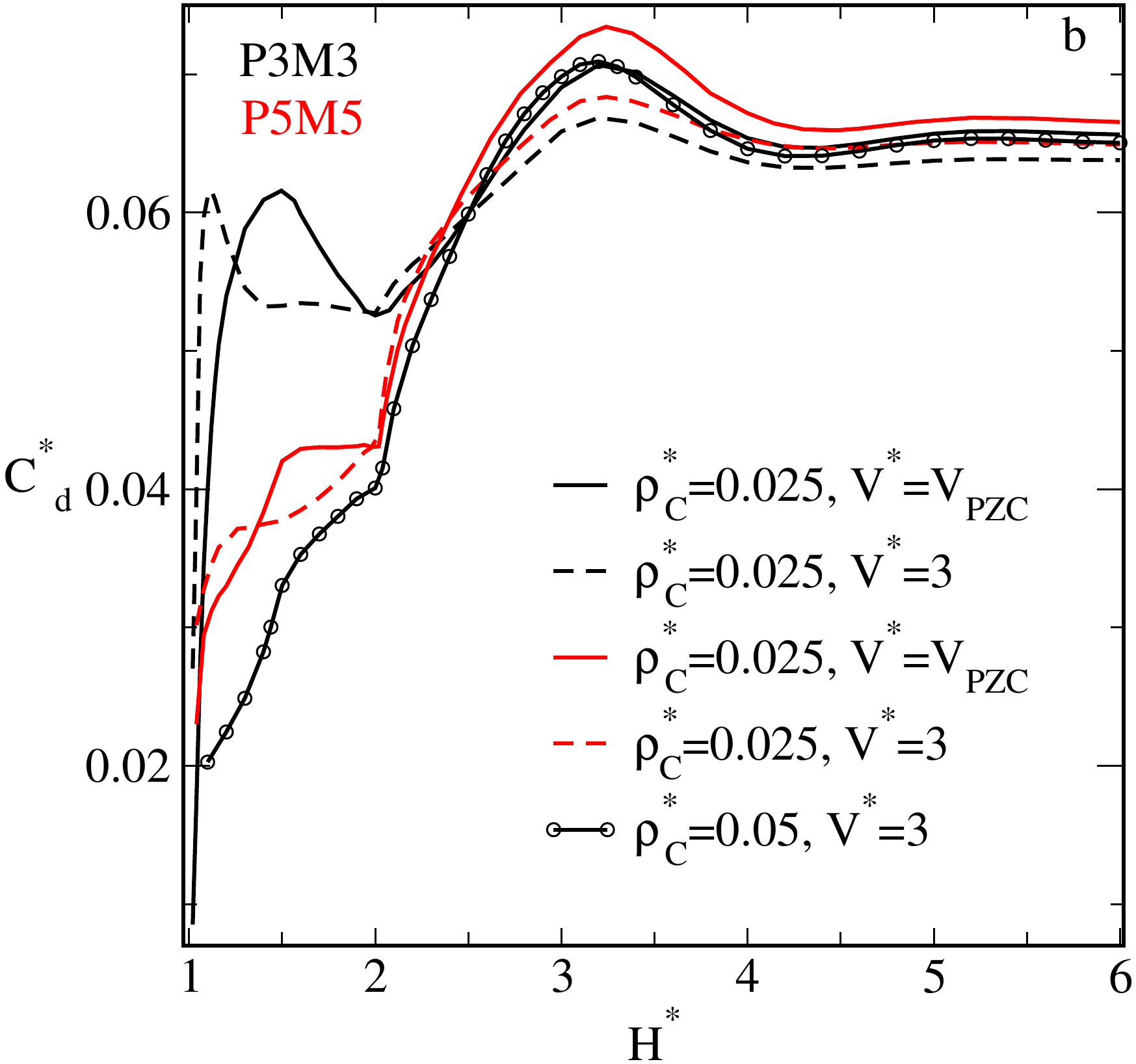}
\end{center}
\caption{(Color online) (a): The dependence of $C^*_\textrm{d}$
on the slit-like pore width, the pore walls are
modified by uncharged chains (N6), $V^*=0.5$.
The parameters are: $\rho^*_\textrm{b}=0.3$, $T^*=0.15$.
The curves are for grafting density
$\rho^*_\textrm{C}=0$,~=-- black curve and circles,
0.05~--- red curve and squares, 0.1~--- blue curve and triangles.
(b): The dependence of $C_\textrm{d}^*$ on the pore width.
The line decorated with circles is for $\rho^*_\textrm{C}=0.05$,
all remaining lines~--- for  $\rho^*_\textrm{C}=0.025$. The calculations are
for the brushes P3M3 and P5M5 and the line nomenclature is explained in
the figure.
Reprinted with permission from reference~\cite{pizio_inpress}. Copyright AIP.
}
\protect
\label{fig9}
\end{figure}

One of the properties of interest  is the differential
capacitance. It is defined as the derivative of the surface charge
with respect to  the applied electrostatic potential, $V^*$,
\begin{equation}
C^*_\textrm{d}=\frac{\rd Q^*}{\rd V^*}\,.
\end{equation}
Recently there has been much interest in the dependence of a differential
capacitance on the pore width
\cite{wu_nano,bazant,feng,gogotsi1,gogotsi2}.
This interest has been inspired by a possibility of developing
new electric double layer capacitors.
In particular,
it is quite challenging for the theory to provide interpretation
of experimental observations concerning an anomalous growth
of capacitance with a decreasing pore width for very narrow
pores.
 A few examples of the dependence of the reduced differential
capacitance on pore width are given in figure~\ref{fig9}.

In the absence of tethered chains, the dependence of $C^*_\textrm{d}$ on the reduced
pore width, $H^*=H/\sigma$, exhibits oscillations, with two well pronounced
maxima at $H^* \approx 1.3$ and 3.3, ($V^*=0.5$) figure~\ref{fig9}~(a).
The maxima are separated by a trough.
In the presence of tethered chains  made of six
uncharged segments (this system is abbreviated as N6), the magnitude
of oscillations decreases,
but at a low grafting density, $\rho^*_\textrm{C}=0.05$,
the capacitance curve still preserves the shape
similar to the case without chains.
At higher grafting density, $\rho^*_\textrm{C}=0.1$, the first maximum is washed out
and only the reminiscence of the second maximum
preserves.

Even a relatively small amount of chains substantially alters
the electric capacitance in narrow pores compared to a pore with bare
walls. Since the segments are uncharged, the observed effect of chains
is due to excluded volume effects.
Figure~\ref{fig9}~(b) shows the results for the pores modified by grafted
polyampholytes, i.e., copolymers consisting of
positively and negatively charged segments
(in some models uncharged groups are considered as well, see e.g.,~\cite{dobrynin1}).
The segments
with charges of different sign
can be distributed randomly or can form certain patterns.
In particular, the charges of the same sign can
be arranged in blocks. The results displayed here
have been obtained for the chains whose net charge is zero,
i.e., the number of positively and negatively charged segments
within a chain is the same.
Two polyampholyte chains were considered~\cite{pizio_inpress}, namely built
of 6 and of 10 segments, whose 3 or 5 initial segments
bear positive unit charges, while the following segments
bear negative unit charges. These systems are abbreviated as P3M3 and P5M5, respectively.
The majority of the
results in figure~\ref{fig9}~(b) are for a low grafting density, $\rho^*_\textrm{C}=0.025$. Higher
grafting density suppresses oscillations, similarly to the
case of uncharged segments.
The electrostatic potential at the walls is low because
pronounced  oscillations appear only at low voltages, as in the case of
non-modified pore walls~\cite{pizio1}. Dashed lines in figure~\ref{fig9}~(b) were obtained
at  $V^*=3$. Solid lines, however, were evaluated
at $V^*=V_\textrm{PZC}$, the potential of zero charge. Note that
for the system involving polyampholytes,
the value of the potential of zero charge differs from zero,
even  for  chain particles whose net charge is zero~\cite{borowko6}.
Of course, the value of $V_\textrm{PZC}$ changes with the pore width.
In the case of non-modified pores, the first (and the most significant)
maximum appears at the distance close to $H^*\approx 1.3$.
The presence of a brush significantly diminishes this maximum. The maximum still
preserves for the brush P3M3, but vanishes for the  brush of longer chains, P5M5.

In the case of a pore with non-modified walls~\cite{pizio5}, the
the distribution of charges inside the pores,
and, consequently, the dependence of $Q^*$ on $V^*$, results from the possibility of
the development of a number of layers of ions at both walls
and from the interference between the layers formed at the opposing walls.
For pores having walls covered
with brushes  whose segments are charged, there
exists a constant amount of charges inside the pore, due to
charged segments. Their distribution is not ``free'', but
is restricted by the existing bonds between the segments.
This ``background'' charge distribution
reduces the freedom of changes of
the distribution of ions and, consequently, variations of $Q^*$ with $V^*$
become smaller.
This is reflected by a smaller (in comparison to non-modified pores)
amplitude of the function $C^*_\textrm{d}$ vs. $H^*$.

Another interesting question is the effect of the
brush on the dependence of the double layer capacitance
on the voltage (or the charge) at the wall. It is known
that for non-modified surfaces, the shape of this
curve changes from a single-hump (``bell-like'') to
a double hump (``camel like'') if the bulk density changes.
For non-modified surfaces, the appearance of single, as well as double humps was found
in Monte Carlo simulations by Fedorov et al.~\cite{fedorov1,fedorov2}, and in
DF calculations by Wu et al.~\cite{wu111} and \cite{henderson1}.
This behavior was  experimentally observed  in the systems that
involve surfaces free from grafted species~\cite{lockett,alam}.

The calculations carried out in reference~\cite{pizio_inpress}
have revealed that for narrow pores modified with grafted polyampholytes,
 even three humps can appear on the dependence of the capacitance on
 the electrostatic potential on the wall and that the shape of the
 capacitance curve depends on the grafting
density.
In general, in the presence of tethered species in  slit-like pores, the shape of $C^*_\textrm{d}(Q^*)$ changes
from a camel-like to a bell-like, if the bulk RPM fluid density changes from low
to moderate and high. The value of crossover bulk density depends on the grafting density of the chains, their length and on the charge sequence along the chains. It is possible to alter
the shape of $C^*_\textrm{d}(Q^*)$ curve from double to single hump by changing
the grafting density, if the bulk fluid density is adequately chosen. The symmetry of
$C^*_\textrm{d}(Q^*)$ curves with respect to $Q^*=0$, observed for several model brushes of uncharged
segments, is destroyed if one explores the brushes made of the chains having charged segments.

To summarize, these few examples show the capability of the DF
approaches to describe the models of complex systems comprising a solid surface with
end-tethered polyelectrolytes or polyampholytes and electrolyte solutions.
There is much room for improving the modelling and theory.
One of the principal problems is an adequate description of the aqueous solvent and
its dielectric properties, as well as to reasonably capture the screening of ion-solvent interactions.
The inclusion of orientation degrees of freedom of species into the DFTs
is, therefore, necessary. In general, we expect that the systems described above will find their place
in the research aimed at optimizing the energy density of supercapacitors, see
e.g.,~\cite{kondrat_s}.

\section{Computer simulations}\label{sec7}

Computer simulation is considered as an intermediate method
bridging up the experimental and theoretical studies.
Theoretical DF approaches have already been discussed above. Concerning
the experimental studies, one should note that  recent technological
progress in the fabrication of modified surfaces involving tethered
chains is not adequately followed by the advances in the methods
for their experimental analysis~\cite{l_patra}. The existing
techniques such as optical, X-ray and neutron scattering, operate
on the meso- and micro-length scales
and cannot be
informative enough for the systems rapidly changing on the nano-scale
(as in the case of nano-patterned surfaces). On the other hand, the
Atomic Force Microscopy (AFM) can be invasive~\cite{l_wilder}, and
can lead to
non-negligible changes of the material structure~\cite{l_patra}.
Other difficulties are connected with a precise control of polymer
synthesis (problems with polydispersity, defects, purity, etc.).

The benefit of computer simulations is that they provide  access to
the positions (and in some methods to the velocities)
of individual particles enabling a high resolution insight into the
structure and internal dynamics of the particles constituting the material. As a result, the
relevant microscopic mechanisms of particular phenomena can be
clarified or even predicted in the way inaccessible for theory
and the experiment. This level of resolution is achieved at
the expense of simplification of the interaction potentials (often being
empirical) and of a finite size of the simulated system.
Most simulations of the systems involving tethered chains can be
split into two categories \cite{booka2,binder_1,rev_comp_sim}:
the (semi-)atomistic (the particle represents an atom or a group
of atoms involving hydrogens, e.g., CH$_2$) and mesoscopic (the particle represents a fragment
of a polymer chain or a collection of solvent molecules).
Both approaches can be thought of as mutually complementary. While the (semi-) atomistic
modelling allows one to address the real chemistry of the system (and the effects
like hydrogen bonding, entanglements etc.), it lacks a possibility to cover large
length- and time-scales of the phenomena under study. The mesoscopic modelling
is capable of doing that, but at the expense of introducing a certain level of
coarse-graining focused on chemical (in)compatibility of fragments and the effects
of their microphase separation. For further details of various particle-based
simulation techniques for the systems involving tethered chains see the reviews
\cite{booka2,binder_1,rev_comp_sim,bind-milch,bind-soft}.

\subsection{Mesoscopic modelling}\label{sec7a}

The mesoscopic level simulations have been mainly based on the
Brownian~\cite{BD1,BD2,BD3,BD4,BD5} and dissipative particle dynamics
(DPD) \cite{DPD_A,l_DPD,l_frenkel,pivkin_a}
methods and we will restrict our review to the latter approach only.

DPD is an off-lattice mesoscopic simulation technique which involves a set of
particles moving in continuous space on a discrete time scale.
Particles (beads) represent whole molecules, fragments of the polymer
chain or fluid regions, rather than single atoms, and atomistic
details are not considered. In other words, the internal
degrees of freedom of particles are integrated out and instead of real
interparticle interactions one introduces effective pairwise forces that
depend on bead-bead distances.
The total force acting on $i$-th bead from the interaction with all other
$j$-th particles is given as a sum
\begin{equation}
 f_i=\sum_{i\ne j} \left[F^\textrm{C}_{ij}+F^\textrm{D}_{ij}+F^\textrm{R}_{ij}\right],
\label{eq:force}
\end{equation}
where $F^\textrm{C}_{ij}$ is a conservative force that  identifies
polymer beads, while the dissipative, $F^\textrm{D}_{ij}$, and random,
$F^\textrm{R}_{ij}$, forces form a thermostat that keeps the temperature of the
system constant.  All the details of the method (e.g., justification of
introduction of effective forces and explanation how to derive them)
are given elsewhere~\cite{DPD_A,l_DPD,l_frenkel}. A key property of
the forces in DPD simulation is that they conserve the momentum locally,
so that hydrodynamic modes of the fluid emerge even for small particle
numbers.

The parametrization of conservative forces is density and temperature dependent
and can be linked to the Flory-Huggins parameter for the case of
mixtures (see, e.g., reference~\cite{groot_a}).
The gain of such a way of coarse-graining is twofold. Firstly,
the number of moving entities is essentially reduced. Secondly, the effective
forces in DPD are much ``softer'' than the atom-atom ones.  This
allows one for using much larger timesteps in the DPD simulation, in
comparison with (semi-)atomistic MD  discussed below. The conservation
of total momentum ensures a correct hydrodynamic behavior of
the systems (i.e., the system satisfies the Langevin hydrodynamic
equations). As a result, the method permits to access much longer time
and length scales and ensures a faster micro-phase separation of
non-miscible species, as compared with conventional MD simulations.
Simulations of polymeric fluids in volumes up to 100~nm in linear
dimension for tens of microseconds are now common~\cite{l_DPD}.

The DPD modelling of confined fluids, or systems with walls, requires
a special care. As it was discussed by Visser et al.~\cite{l_visser}, the
central problem is how to construct a solid wall for mesoscopic
modelling.  At this level of modelling, three boundary conditions must
hold for a solid wall: (i) impenetrability (particles are not allowed
to enter the wall), (ii) the wall should not artificially affect (i.e.,
by imposing an artificial structure) the fluid properties, and,
in the case of flows, (iii) no-slip; the wall should
impose velocities in accordance with Langevin hydrodynamics. Previous
studies succeed to implement impenetrability and no-slip boundary
conditions. They usually used several layers of frozen particles residing on a
surface~\cite{l_tilde,l_1,l_2,l_4} to ensure impenetrability. This, however, induces the
ordering near a surface, which propagates into a substantial portion of
the system. This effect has a physical justification for the atom-based simulation when
one mimics the crystal-like structure of the wall, but should be
interpreted as artificial for the meso-scale type of modelling.

Quite recently, a new approach was introduced for modelling the walls in DPD studies
\cite{l_ilny}. It was assumed that, if the $i$-th bead
with coordinates $\mathbf{r}_i = \{x_i , y_i , z_i \}$ approaches the
surface, e.g., the plane $z = 0$, it interacts with an ``imaginary''
bead that has the coordinates $\mathbf{r}_j = \{x_i , y_i , 0\}$.  The
total wall-bead interaction has all three contributions given by
equation~(\ref{eq:force}).  Repulsion of the surface can be tuned by
adjusting the parameters of conservative force, $F^\textrm{C}$.  The condition
of total momentum conservation for the wall-bead interaction is
applied by the following procedure. The force acquired by each wall
due to the interaction with the approaching beads is accumulated during
the evaluation of forces. Then, it is evenly distributed over all the beads
inside the simulation box.  The effective surface, therefore, is
structureless and, thus, it does not introduce any artificial
perturbation of the fluid structure near the wall.  The details
concerning DPD simulations can be found in
references~\cite{l_DPD,l_tilde,l_ilny,l_ilny2,l_groot,l_tilde1,l_tilde2}.

Let us now briefly summarize the most interesting DPD results for the
systems involving tethered polymers. One of the first DPD studies of
polymer-coated surfaces was carried out by Gibson et al.~\cite{l_gibb},
who investigated the adsorption of colloidal particles on a surface
modified with pinned polymers.  In particular, it was found that the
adsorption of colloidal particles was smaller when the size of the
polymer relative to the colloidal particle and the density of the polymers
increased. Moreover, the adsorption reduces when the polymer is well
solvated.

Malfreyt and Tildesley~\cite{l_tilde1} simulated the brushes
tethered at two walls of a slit in a good solvent.  The density
profiles of polymer segments across the pore were found to exhibit
parabolic shape, and diffusion along the pore axis was significantly
greater for the solvent particles in the middle of the slab than in
the polymer region, since the solvent molecules were trapped within the
entangled polymer.  Later, Goujon, Malfreyt, and
Tildesley~\cite{l_tilde} extended DPD method to the grand canonical
ensemble and studied the compression of grafted polymer brushes in a good
solvent.  They also studied the compression of polymer brushes under shear
and evaluated the friction coefficient as a function of compression,
shear rate and properties of the solvent~\cite{l_tilde2}. The
constant and oscillatory flow of a fluid between two plates with
tethered chains was simulated by Wijmans and Smit~\cite{l_wijamans}.

Pastorino et al.~\cite{l_pastorino} investigated a short-chain melt
between two brush-covered surfaces.  The equilibrium situation, as
well as the one under shear flow were explored.  The polymers of both brush
and melt were assumed identical.  They studied the interdigitation of
the melt and the brush and also considered the orientation of bond
vectors on different length scales, as well as radii of gyration,
end-to-end vectors of free and grafted chains, and in the case of flow
simulations, the velocity profiles. In the second work, Pastorino et
al.~\cite{l_pastorino1} used the Gibbs criterion to localize the
brush-melt interface and analyzed its equilibrium fluctuations in
terms of a capillary wave Hamiltonian augmented by an elastic term
that accounts for the deformability of the brush.  However, one should
stress that the simulation method used by Pastorino et
al.~\cite{l_pastorino,l_pastorino1} was different from the standard DPD
scheme~\cite{l_frenkel}. Namely, they assumed that the interactions
between the polymer segments and polymer segments with surfaces were
of the Lennard-Jones type, and they performed molecular dynamics
simulations with DPD thermostat.

The effects of chain architecture on the structure of tethered
rod-coil polymer layers was studied by Li et al.~\cite{l_li}. When
immersed in a selective solvent for the coil blocks, rod blocks tend
to form aggregates. Linear and Y-shaped polymers exhibited a similar
aggregative behavior, but comb-like brushes were found to possess more
diverse aggregative behavior compared to linear brushes.  Surface
structures with aggregates taking form of cones, cylinders, or layers
of spheres were found. The behavior of grafted binary polymer brushes
with compatible components in the case of different chain lengths was
investigated by Xue et al.~\cite{l_xue}. They observed layered
structures parallel to the surface indicating ``phase separation'':
short chains were suppressed in the layer adjacent to the surface,
whereas longer chains were to much extent stretched.  By slightly
changing the solvent selectivity in preference of shorter chains, inversion
of the layered structure was found.

\begin{figure}[htb]
\begin{center}
\includegraphics[width=12.0cm,clip]{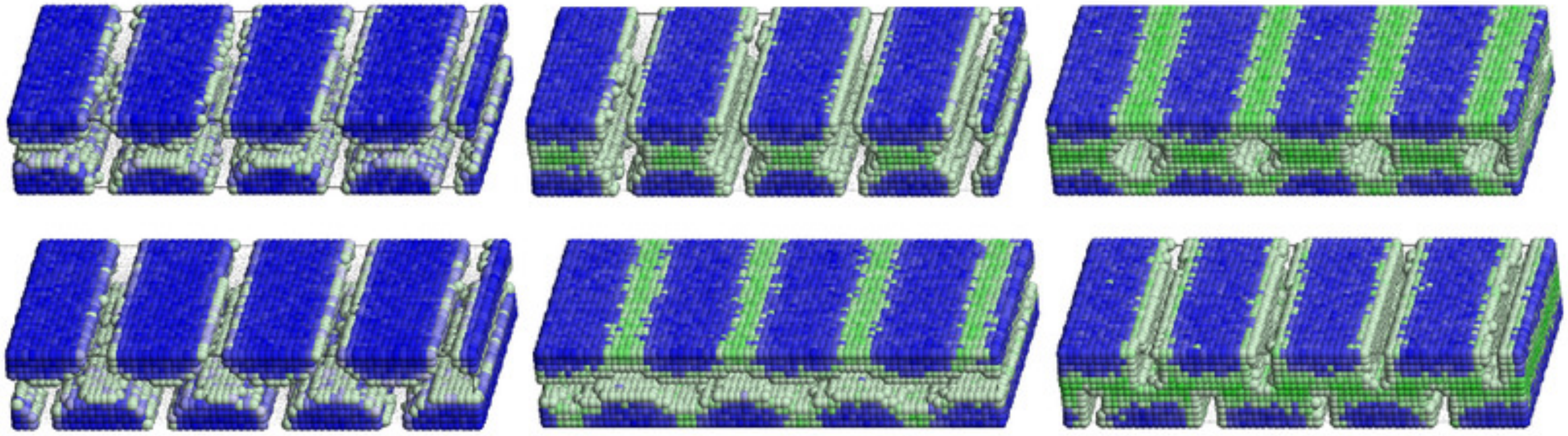}
\end{center}
\caption{(Color online) Solvent-mediated morphologies formed by the polymer brush
  and the solvent mixture with an increase of the good solvent
  fraction (from left to right). Two cases are shown of the in- and
  out-of-phase arrangement of stripes, top and bottom rows of images,
  respectively. The plot is density color-coded showing the
  polymer-rich domains (in blue) and good solvent-rich domains (in
  green). The ``empty'' space is occupied by a ``bad'' solvent. One should
  note solvent-mediated switching between lamellar and ``meander''-like
  morphology in the bottom row of images. Reprinted with permission from
  reference~\cite{l_ilny2}. Copyright CMP.}
\protect
\label{iln_sok}
\end{figure}

There have also been performed several DPD studies of the brushes tethered
to non-uniform surfaces, see
e.g.,~\cite{l_patra,l_ilny,l_ilny2,petrus1,petrus2}. The systems
involving brushes attached to some parts of the surface, the so-called
structured brushes, are potentially very important in the manipulation
of fluids at very short length scales.  A prototype case consists of a
solvent in contact with a surface, or confined between two identical
plane-parallel substrates, decorated with stripes of tethered chains
that alternate periodically in one direction (say, $X$) and are
infinite in the other transverse direction ($Y$). The performed
simulations indicate that heterogeneity of tethered layers has a great
impact on the structure of confined fluid and on thermodynamic and
dynamic properties of the systems.  In particular, Ilnytskyi et
al.~\cite{l_ilny} employed DPD to investigate the behavior of a binary
mixture of A and B, exhibiting a demixing in a bulk phase, confined in
slit-like pores with walls modified by the stripes of tethered brush
of chains made of beads A. The main issue was to determine possible
morphologies that can be formed inside the pore depending on the
geometrical parameters characterizing the system (the size of the pore
and the width of the stripes), cf. figure~\ref{iln_sok}.  In order to describe the observed
morphologies, these authors calculated several characteristics, such as the
density and local temperature profiles, the radii of gyration for the
attached polymers, and the minimum polymer-polymer distances in the
direction parallel and perpendicular to the pore walls. The summary of
findings was presented as a sketch of the diagram of morphologies.

The study of reference~\cite{l_ilny} has been extended  to the case of different (so-called
in- and out-of-phase) arrangements of the stripes at two surfaces and
varying composition of the confined mixture~\cite{l_ilny2}. For
in-phase narrow stripes and narrow pores, the case is reduced
to a quasi two-dimensional micro-phase separation within the mixture confined
between the polymer-rich lamellae on each wall. On the other hand,
for narrow pores and moderate stripe widths,
the in-pillar
separation is
quasi one-dimensional. In the case of wider pores and very wide stripes, the
pore interior splits into the regions showing quasi two-dimensional and bulk-like
behavior. A number of solvent-mediated morphologies are observed when
the stripes are weakly separated, both along the wall and across the pore.
With an increase of
the concentration of A beads in the mixture, they  bridge up
the striped brush in either or in both directions. Certain effects of
morphology switching are observed. On the application side, the
concentration driven morphologies can be fixed permanently via
crosslinking, if a mixture contains a short crosslinker compatible
with the polymer. Fixed morphology could be used as a nanotemplate or
as a microfluid channel~\cite{l_ilny2}.

An interesting work on the application of DPD simulation technique to
describe brushes on nanopatterned surfaces was published by
Patra and Linse~\cite{l_patra}.  In the model considered by them,
polymer chains were randomly attached onto an infinite long stripe
along the $Y$ axis with finite width $\Delta$ along the $OX$ axis. The
polymers were treated as jointed chains composed of $M$
spherical subunits (beads) connected by bonds. Each bead corresponded
to a small part of a real polymer, approximately one Kuhn
length~\cite{l_booka3}, and was described by a soft repulsive
potential.  The simulations were carried out for a range of widths of
the stripe, $\Delta$, for several grafting densities $\rho_\textrm{C}$ and for
several numbers of beads, $M$.

\begin{figure}[!b]
\begin{center}
\includegraphics[width=0.98\textwidth]{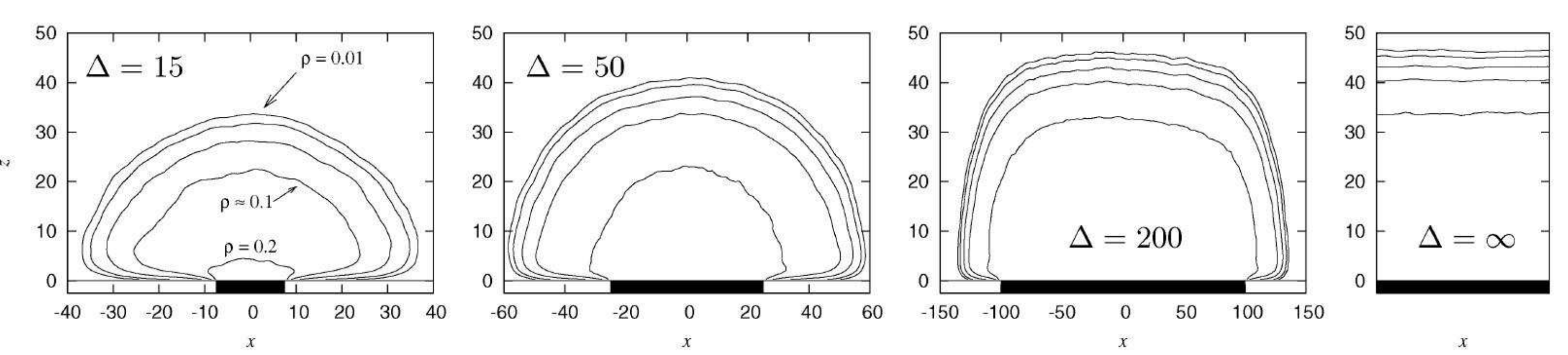}
\end{center}
\caption{The contours of the beads density
for polymer length $M=100$ and grafting density
$\rho_\textrm{c}=0.11$ at stripe width $\Delta=15,\, 50,\, 200$, and
$\infty$ (from left to right). The contour curves mark the densities 0.01, 0.02,
0.05, 0.1, and 0.2. The location of stripes is indicated
by the filled black areas. Note, that the scale of the abscissas differs among the panels.
Reprinted with permission from reference~\cite{l_patra}. Copyright ACS.}
\protect
\label{fig_con}
\end{figure}

In figure~\ref{fig_con} we show examples of the local densities of the beads
as functions of $X$ (along the surface) and $z$ (in the direction
perpendicular to the surface). The calculations were carried out for several
values of $\Delta$ (all distances are measured in the unit of
the bead diameter).

As $\Delta$  is increasing, the following
trends in the system were observed: (i) an increase in the
brush height $h$, (ii) an expansion of the region with high density
close to the grafting surfaces, and (iii) an appearance of a
considerable overshot of polymers outside the grafting region.
The polymer density fell off rapidly when the perimeter of the brush was reached.

\begin{figure}[htb]
\begin{center}
\includegraphics[width=8.0cm,clip]{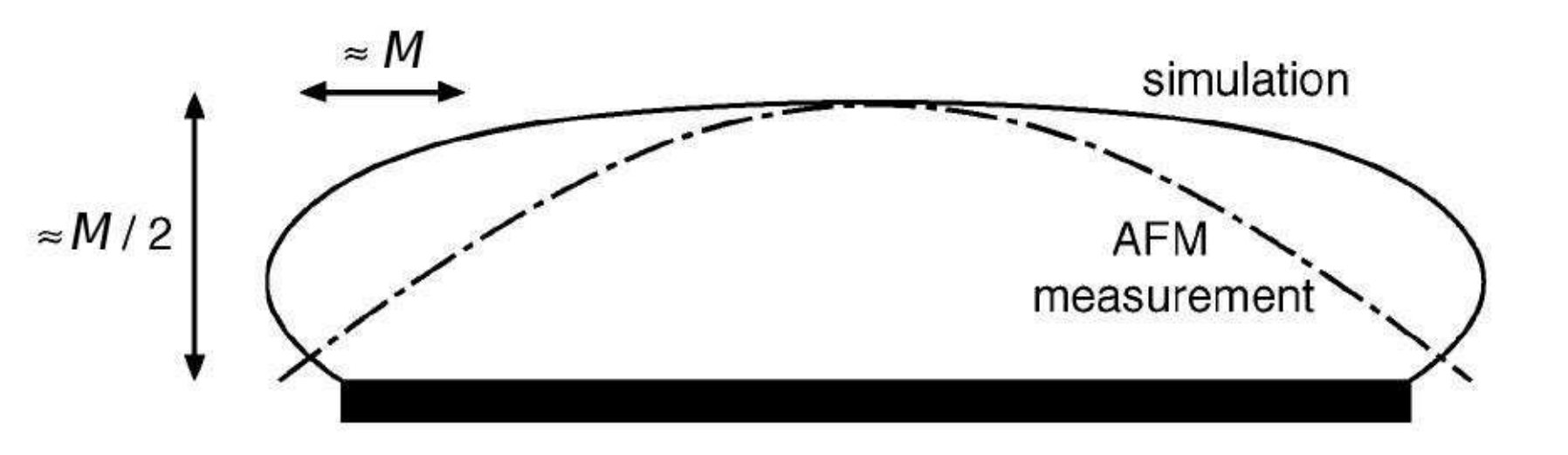}
\end{center}
\caption{Typical shapes of a nanopatterned brush as computed
by simulation (solid line) and as measured by AFM (broken line).
Horizontal and vertical length scales have been indicated in the
figure. The location of the stripe is indicated
by the filled black area. Reprinted with permission from reference~\cite{l_patra}. Copyright ACS.}
\protect
\label{fig_con1}
\end{figure}

Formally, the brush profile over a stripe can be evaluated by using
AFM. However, the AFM tip, when approaching the surface, deforms the brush
and, therefore, the height obtained from experiments differs from
that for undeformed brush.
Typical shapes of brush profiles from simulations
and AFM measurements, as estimated by Patra and Linse~\cite{l_patra},
are displayed  in figure~\ref{fig_con1}. In contrast
to the simulated profile, the experimental one does not display
a clear plateau in the center of the stripe even though the brush
is $10\div 100$ times as wide as it is high. Patra and Linse
also present a strict derivation for the maximal decrease of the
brush height outside  the center. Since a polymer brush is a soft material, it bends under the
force of the AFM tip. This behavior was also obtained by MD
simulations of homogeneously grafted polymer brushes~\cite{l_murat}.

The simulations by Patra and Linse have been carried out
in the domain $M\leqslant \Delta \leqslant 5M$. Their results
suggest that this domain contains the
transition from narrow polymer brushes to the regime of wide
brushes. An improved understanding of this
transition calls for a reassessment of the experimental
conditions of the AFT technique, but even
more exciting task would be to investigate the experimentally
preferable orientation and conformational changes of chains
near an edge by using, for example, labeled polymers.
Many applications of nanopatterned polymer brushes would benefit
from such a deeper understanding of the molecular arrangement
in narrow polymer brushes.

\subsection{Atomistic and semi-atomistic modelling}\label{sec7b}

The strength of mesoscopic simulation methods, discussed in the previous
subsection, is focused around their capability to cover nano-scale structure
and the dynamics of polymer brush and the (mixture of) fluids contained within the pores.
However, the soft interaction potentials, used in such studies, are typically
of heuristic form and their relation to the real chemical substances is still
an open question. Rigorous coarse-graining methods are still to be
 developed \cite{multiscale1,multiscale2,multiscale3,multiscale4}.
Therefore, the atom-based simulations have a twofold purpose: (i) to be used on smaller
scale systems to serve as a feed for the estimates of coarse-grained interaction
(so-called multiscale approach) and (ii) to be an independent tool for more ``chemically
detailed'' simulation studies.
An extensive literature is available on the applications of \mbox{(semi-)atomistic}
models, reviewed earlier in references~\cite{bind-milch,bind-soft,l_mila,l_simul2,l_simul4}.
Here, we outline only some of the works that are of interest or
have been omitted in previous reviews.

One of the major tasks of computer simulations is to verify the analytic
predictions from the SCF theory for grafted polymer layers.
The MC simulation using the bond-fluctuation model was employed in \cite{nowe-1}
where the quantities describing the equilibrium structure of the brush are derived from the SCF
theory and compared with the MC data without free parameter. In most cases, the
results are in agreement with the SCF predictions. The causes of discrepancies are also discussed.
The MD simulation of polymeric brushes in solvents of varying quality is presented
in reference~\cite{grest}
where the resulting equilibrium structures are compared with the predictions of SCF approaches.
The scaling of the brush height was found to agree with SCF
predictions at temperatures $T>T_\theta$, where
$T_\theta$ is the $\theta$ temperature \cite{gennes1} of the solvent.
For $T<T_\theta$, the agreement was poor, which was connected with
the separation of the brush into solvent-rich and solvent-poor phases.
This phase separation
was found to disappear with an increasing grafting density and chain length.
The effect of a weak attractive
surface interaction on the structure of the brush was also studied.

The entropy of polymer brushes in a good solvent limit was evaluated numerically
in reference~\cite{l_si2}.
The cases of a free polymer brush, a compressed polymer
brush and two polymer brushes facing
each other were studied. The total number of configurations, entropy, and
the force acting on the polymer
were calculated for various
chain lengths and grafting densities by means of MC simulations
using an efficient enrichment algorithm
on a simple cubic lattice. It was demonstrated that the effect of an excluded volume plays an
essential role and that the entropy of a free polymer brush drops as
(spacing)$^{-2.4}$,
while the entropy of a compressed polymer brush and two polymer brushes agrees very well with
the analytical SCF theory by Milner, Witten, and Cates \cite{milner_2610}.
The analysis has been also extended to the case of a polymer solvent.

A MD simulation of polymeric brushes immersed in a melt of mobile polymer
chains was presented \cite{grest_5532}.
The brush height and segment density profiles were evaluated for chains of
length $M$ grafted at one end to a solid surface, immersed in a melt of
polymers of the same type. As
the chain length $P$ of free chains increases, there is a crossover from a wet to a dry brush in
agreement with scaling and SCF theories. Since the interaction between all
segments is identical, this crossover is driven purely by entropy. The simulation
results were compared with earlier simulations.

The exchange of chains between the brush and the bulk solution
is of obvious relevance in understanding the thermodynamics
and the kinetics of the formation of a grafted layer. In reference~\cite{lai_669},
by including a finite head-group adsorption energy,
the bond-fluctuation model has been used to simulate the polymer brush in a
good solvent with the effect of chain exchange between the brush and the
bulk. The self-adjusted surface coverage was measured for different values
of chain lengths and head-group energies. It was found that
the polymer chains in a grafted layer are replaced by introducing shorter chains of
identical head groups, which supports some experimental studies.
The irreversible adsorption of single chains grafted with one end to the surface was
also studied by combining the scaling theory with computer simulations in reference~\cite{descas}.
Using scaling arguments, the authors derived relations between the
overall time of adsorption the characteristic time of adsorption, and the chain
length. To support the analysis, the MC simulations were performed using the bond
fluctuation model. The sequence of adsorption events is reproduced very well
by simulations, and an analysis of various density profiles
supports the theoretical model.

To clarify the differences in the properties of grafted chains at various
grafting densities, a number of simulations were undertaken.
In particular, the MC simulations were performed using bonded Lennard-Jones spheres.
These simulations were
aimed at observing the mushroom-to-brush transition at low grafting densities
as well as the properties of a high density brush \cite{l_si3}.
Through approximate scaling arguments, the location of the mushroom-to-brush transition
was estimated and found to agree closely with the grafting
density at which no more chains with mushroom dimensions
can be inserted without overlap. Comparisons with analytical SCF
theory for poor solvents were found to be much less favorable.
In this case, the density profiles substantially differ
from the SCF picture due to high segment densities and
cluster formation. In reference \cite{he_6721},
the static and dynamic properties of polymer brushes of moderate and high
grafting densities were studied using MD simulations and the Langevin-type model.
The fluctuation dynamics in the direction lateral to the surface was found to be
well described by Rouse scaling. The dependence of the dynamics on the grafting
density was well described by the dynamic behavior of thermal blobs
for fluctuations in the direction both perpendicular and parallel to the surface.
The pulling forces acting on the polymer bonds were also studied.

\begin{figure}[htb]
\begin{center}
\includegraphics[width=8.0cm,clip]{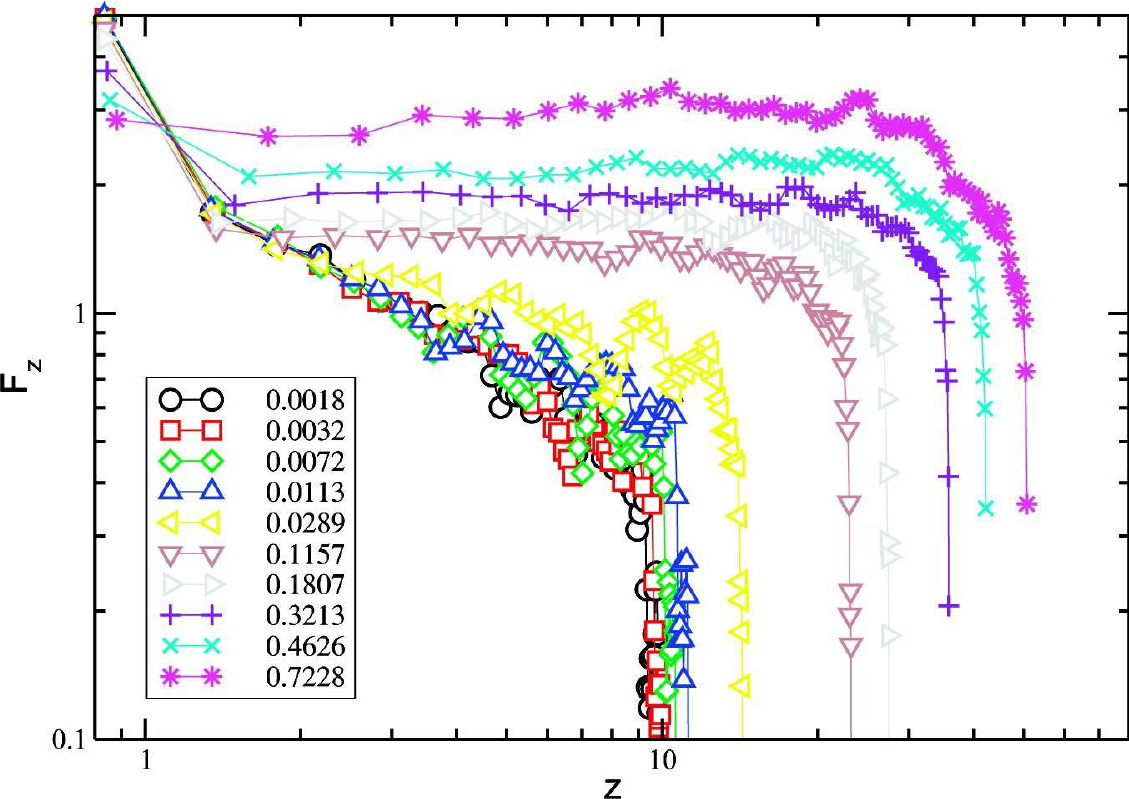}
\end{center}
\caption{(Color online) Averaged bond forces along the $z$-direction for $M=64$ as
a function of the distance from the surface
at different grafting densities, given in the figure.
 Reprinted with permission from reference~\cite{he_6721}. Copyright ACS.}
\protect
\label{pulling}
\end{figure}

The knowledge of the  dependence of the bond forces on the grafting density is important
to develop strategies for making brushes with high grafting densities.
The MD method allowed the authors of reference~\cite{he_6721}
 to calculate the average vertical forces pulling
the chains and the corresponding grafting energies.
The binding potential between the nearest chains was the FENE potential,
for details see reference~\cite{he_6721}.
 The obtained pulling force
as a function of the average distance from the surface
for chains of $M=64$ segments is displayed in figure~\ref{pulling}. The results were calculated
at different grafting densities (shown in the plot).
 At the
top of the brush, the stretching forces vanish, as long as the grafting density
remains moderate. At high grafting densities, even the end bond displays some
residual stretching as a result of strong  interactions
 with neighboring chains. Moreover, at high and moderate grafting densities,
one observes a plateau which indicates the brush regime.

The wetting behavior of polymers on top of a brush consisting of identical polymers
has been studied using the computational methodology developed for a direct
measurement of a wetting transition and its order via the effective interface potential.
The method also permits to estimate the contact angles in the non-wet state and to study the adsorption
isotherms \cite{l_si1}.
In the absence of long-range forces, the system shows a sequence of non-wet, wet, and
non-wet states as the
brush density is increased. Having included attractive long-range interactions, we can make the polymer
wet the brush at all grafting densities, and thus both first- and second-order wetting transitions are
observed. The study highlights a rich wetting behavior when competing adsorbent-substrate interactions of different scales are tuned over a broad
range.

Polyelectrolyte brushes have also been simulated using the MD
method \cite{he_7845}.
These were aimed at studying the fully charged polyelectrolyte brushes with salt,
and the results are compared with SCF theory including finite stretching and
volume effects \cite{bie_6254}.
The SCF approach was found to be capable of reproducing the brush heights at different grafting
densities, salt concentrations and chain lengths
at a semi-quantitative level. At high grafting densities, the density profiles obtained
with both techniques exhibit a particularly close agreement, while at low densities,
systematic deviations between their shapes are observed. The approximation of local
electroneutrality, which the SCF approach is based on, was studied, and its implications
were discussed.

\begin{figure}[htb]
\begin{center}
\includegraphics[width=8.0cm,clip]{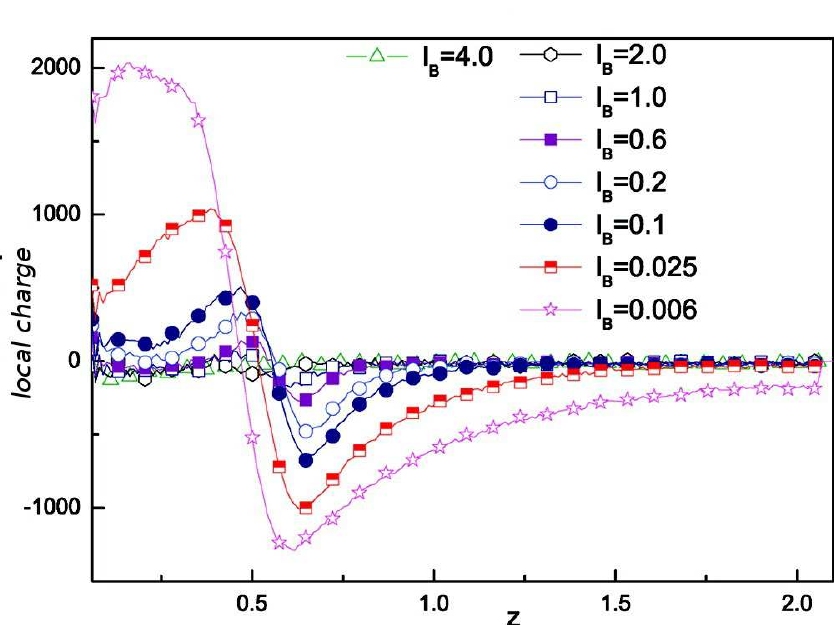}
\end{center}
\caption{(Color online) Local net charge
$q(z)$ as a function of the vertical coordinate for different Bjerrum lengths, given
in the figure.
 Reprinted with permission from reference~\cite{he_7845}. Copyright ACS.}
\protect
\label{polyelevtrolyte}
\end{figure}

It is instructive to compare the local
net charges of the vertical layers of the brush (figure~\ref{polyelevtrolyte}).
This quantity serves for validation of the local electroneutrality assumption
 made in the SCF approach. Not surprisingly, this assumption is satisfied rather well when
the Bjerrum length is large (i.e., for strong electrostatics) and less within the
 weak charge regime. Throughout the osmotic regime (we recall that
in the osmotic regime, the concentration of counterions inside the polyelectrolyte layer
is higher than in the bulk solution \cite{minko1,andra1}),
there exists a
charge separation near the brush surface. The counterions form a charged layer just above
the brush as a result of their osmotic pressure. It is this layer that generates
the osmotic pull which contributes to the chain stretch. On the contrary,
the local charge neutrality of the osmotic regime is rather well satisfied deep
inside the brush. In the case of thick brush layers, when the surface contributes
just a small fraction to the brush height, the approximation of local neutrality would turn
 even more accurate \cite{he_7845}.

One of the topics extensively studied by using computer simulations
is the  adsorption of fluids on the surfaces modified by tethered
chains~\cite{l_si1,l_si2,l_si3} and  the chromatographic process on columns
filled with adsorbents modified by
tethered oligomers~\cite{l_ch1,l_ch2,l_ch3,l_ch4,l_ch5,l_ch6,l_ch7,l_ch8}.
A popular technique in the liquid chromatography
is the Reverse Phase Liquid Chromatography (RPLC).
The mobile phase in RPLC is a mixture of water and
an organic solvent (usually methanol or acetonitrile).
The stationary phase consists of a support material
(usually beads of silica of micrometer size) and a bonded
phase. The bonded phase is most often built of C18 alkyl chains.
The retention of solutes occurs primarily within or on top of the bonded
alkyl chains of the stationary phase.

The mechanism of retention in RPLC has been
of interest for over three decades,
but many of its  aspects are still
highly debated~\cite{l_gri2,l_gri3}.
In general, solute
retention in RPLC
is driven by the free energy gradient on passing
from the mobile phase into the stationary phase.
One of the pioneering MD simulations
of the transfer of a simple nonpolar
solute from a water/methanol solvent mixture into a C18 stationary phase at room temperature
was carried out by Klatte and Beck \cite{klatte}.
These authors, in addition
to a detailed examination of the local environment of the solute, computed
the  excess chemical potential
profiles.
The free energy change was consistent in magnitude with that expected for hydrophobic transfer
from water rich to oil phases, but specific interfacial effects drew
into question the bulk partitioning models.

To elucidate the molecular-level structural
features that control shape-selective separations, the MD
was performed for chromatographic models with alkylsilane length
and temperature analogous to actual materials \cite{lippa}.
Alkyl chain order was found to increase both with an increase of
the chain length and with a reduction of the temperature, as observed
experimentally. Chromatography models of various chain lengths and over a
temperature range that represents highly shape-selective
RPLC stationary phases contained a series of well-defined
rigid cavities.  The size and depth of these ``slots''
increased for the C30 models, which may promote the
enhanced separations of larger size shape-constrained
solutes, such as carotenoids \cite{lippa}.

According to modern theories of RPLC, two main questions regarding the retention
mechanism appear: (i) is solute retention better described by partitioning
into the stationary phase that resembles bulk liquid-liquid
partitioning or by adsorption onto the stationary-phase surface?
(ii) what is the driving force for the retention process, unfavorable
interactions with the mobile phase or favorable interactions with
the stationary phase?
The answer of the above questions
was considered in a series of papers by Siepmann and
co-workers~\cite{l_ch2,l_ch3,l_ch4,l_ch5,l_ch6,l_ch7,l_ch8},
who employed an advanced MC simulation method.
In particular, in an interesting recent work~\cite{l_ch7},
the effects of stationary phase and the solute chain length,
as well as the solute adsorption and separation were studied.
The surface was modified by grafting
dimethyl triacontyl (C30), dimethyl octadecyl (C18),
dimethyl octyl (C8), and trimethyl (C1) to silane surface.
The modified surfaces and the bare
silane surface  were in contact with a
water/methanol (33\%  of methanol) mobile phase that contained
normal alkanes and alcohols.
The simulations were carried out using
 the Gibbs ensemble and the details are given in \cite{l_tildes,l_tildes1,l_ch7}.
\begin{figure}[htb]
\begin{center}
\includegraphics[height=9.0cm]{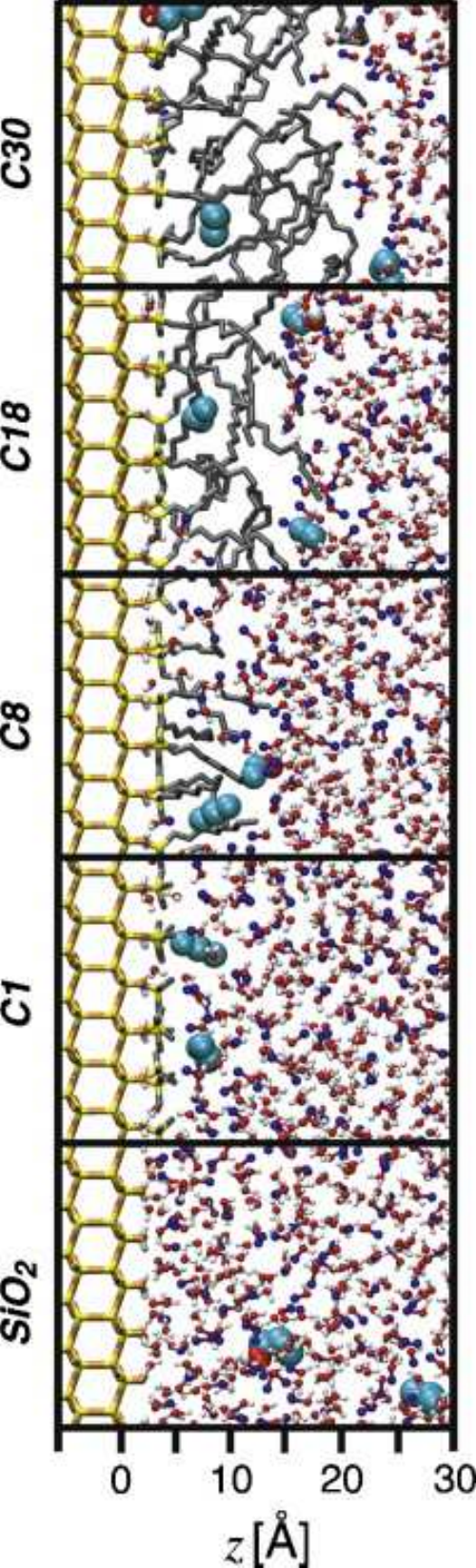}
\includegraphics[height=9.0cm]{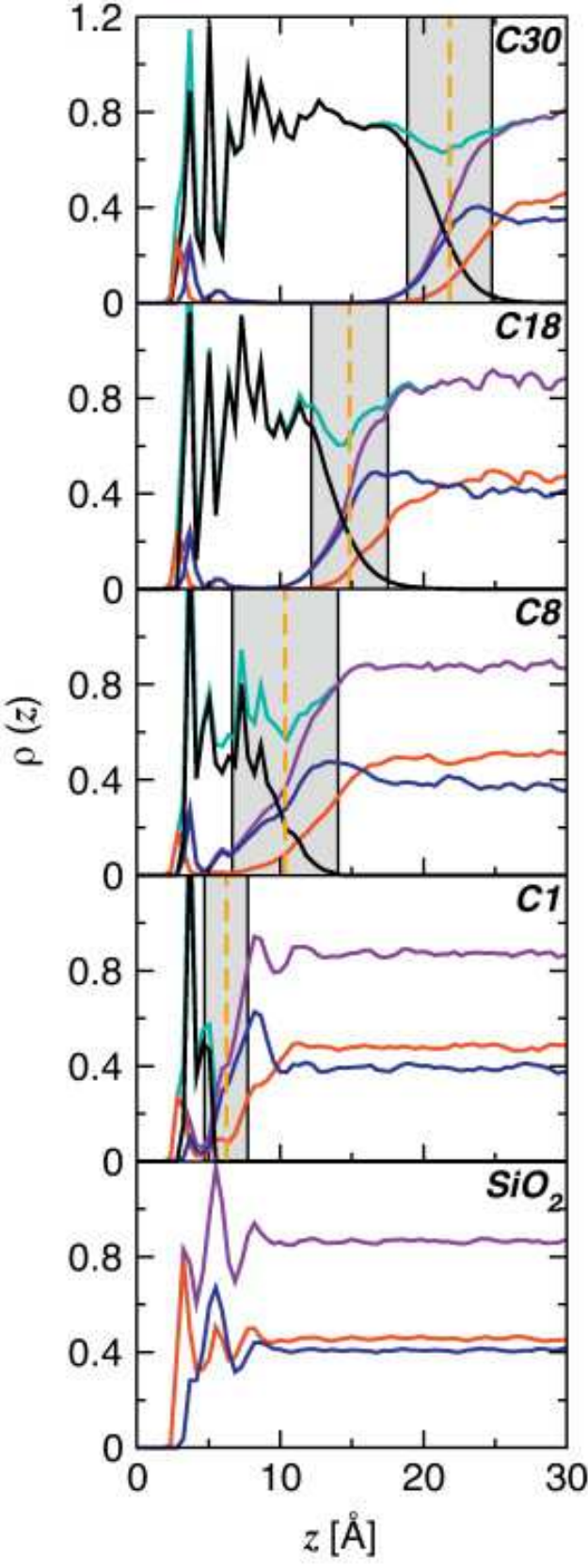}
\end{center}
\caption{(Color online) (a): Examples of the simulation snapshots of the five systems obtained
by Siepmann et al.~\cite{l_ch7}. The stationary phase is shown
as tubes with silicon in yellow, oxygen in orange, hydrogen in white and carbon in
black. The mobile phase is shown in the ball and stick representation with oxygen
in red, hydrogen in white, and carbon in blue. The analyte molecules are depicted
by large spheres with oxygen in red, hydrogen in white, and carbon in cyan.
(b): System composition as a function of distance from the silica surface.
Stationary phase carbon density is shown in black, methanol in
blue, and water in red. The total system and solvent densities are shown in cyan and
purple, respectively. The GDS is shown by the orange dashed line and the interfacial
region is shaded gray. Reprinted with permission from reference~\cite{l_ch7}. Copyright Elsevier.
}
\protect
\label{fig_siep}
\end{figure}

The computer simulations were carried out at
$T=323$~K and pressure of 10~atm, i.e., at the conditions typical of  RPLC
experiments.
Figure~\ref{fig_siep} (panel a) gives typical snapshots of the configurations
of the molecules in the selected systems.

Simulating  tethered molecules,
Siempmann et al. \cite{l_ch2,l_ch3,l_ch4,l_ch5,l_ch6,l_ch7,l_ch8}
investigated how the structure of tethered layer
changed with the length of
the chains.
In  particular, they measured the
dihedral angles, defined as the consecutive angles
between carbons separated by two bonds and the normal to the silica substrate, as
well as the so-called  ``gauche defects''~\cite{ula1} (defined as the
fraction of
dihedral angles in the entire chain deviating by more than 60$^\circ$ from
the angle of the trans-conformer.
They  also studied the local composition and stationary phase thickness.
For this purpose, density profiles for water,
methanol, and stationary phase carbon atoms were evaluated, cf. figure~\ref{fig_siep} (panel b).

The plots also show the Gibbs dividing surface (GDS, i.e., a plane defining
the border between the mobile and stationary phases) and the width of the interfacial region.
The latter  was defined as the range of $z$-values
where the total solvent density lies between 10\% and 90\% of its
bulk value.  However, due to very large oscillations in the total
solvent density, it was not possible to determine a GDS and the interfacial region
for the bare SiO$_2$.
One can see several similarities
between the systems with the chains of different length.
First, one observes a peak of
the methanol local density just above the location of the GDS,
i.e., there is a significant enrichment of the organic
modifier in this region that extends deep into the interior of the
tethered chains. The enrichment  of methanol
at the surface was similar for all lengths of chains.
Second, there is a depletion of the total system density for all
systems (see cyan lines), i.e., a dewetting
effect at the  surface.

However, there are also some differences between the system
with different chains.
 For the C1 system the interface is very sharp
and the methanol enrichment is very pronounced.
The solvent density shows an oscillating behavior
while for other chains it decays smoothly in the interfacial region.
For C8 system,  the interfacial width is larger than for the other
systems, and one observes a more significant solvent penetration in
the C8 system,  as compared to the C18 and C30 stationary
phases. This effect was interpreted by examining the snapshots
in figure~\ref{fig_siep}. The C8 phase
is incapable of completely covering the silica surface, and thus the filaments of
solvent molecules bridging the substrate and the mobile-phase region
prevail. In the case of C18 and C30 phases, they form
a continuous layer with respect to the solvent penetration, and solvent bridges
are extremely rare for  C18 phase. In other words, the C8
phase forms a more heterogeneous surface layer.
It should be stressed that
the water and methanol densities in the interior region of the C30
phase are about one order of magnitude smaller than the corresponding densities
for the C18 phase.

The simulations by Siepmann et al. indicated,
that
the retention of butane
increases with an increasing
chain length, which is in agreement with experimental
data.
The nonpolar alkane solute shows a mixed retention mechanism
with contributions from adsorption in the interfacial region
and partitioning into the stationary phase.
For the C1 phase, retention is dominated by adsorption.
This phase is too thin to allow for partitioning. However,
the  adsorption of butane on a ``hard''
surface of C1 is different from adsorption on a
more flexible, or ``soft'' surfaces involving longer chains.
Siepmann et al. have discussed not only the mechanism of retention
of butane but also other molecules,
and their calculations have clearly demonstrated that advanced
computer simulations can be very helpful in explaining the
mechanism of chromatographic separation.

The absorption (confinement) of free linear chains into a polymer brush  was
studied with respect to chain size $M$ and
the Flory-Huggins type compatibility parameter $\chi$
with the brush by means of MC simulations at various grafting densities using a
bead-spring model. Different concentrations of free chains were examined. Contrary to the
case of $\chi=0$, where all species are almost completely
ejected by the polymer brush irrespective of their length $M$, it was found that for $\chi<0$,
the absorbed amount undergoes a sharp crossover from weak to strong absorption, discriminating
between oligomers, $1\leqslant M\leqslant 8$, and longer chains. For a moderately dense
brush, the longer species, populate predominantly the deep inner part of the brush,
whereas in a dense brush they penetrate into the ``fluffy'' tail of the dense brush only.
Gyration radius $R_\textrm{g}$ and end-to-end distance $R_\textrm{e}$ of absorbed chains thereby scale with length $M$ as free
polymers in the bulk. Three distinct regimes of penetration kinetics of free chains were found regarding the length $M$, in which the time of absorption grows with $M$ at a different rate \cite{milche_aa}.

To summarize this part, one could just cite some simulation works that target technology-related
aspects of polymer brushes. For instance, the authors of reference~\cite{softcond}
remark that ``via computer simulations, we demonstrate how a densely grafted layer
of polymers, a brush,
could be turned into an efficient switch through chemical modification of some of its
end-segments. In this way, a surface coating with reversibly switchable properties can
be constructed. We analyze the fundamental physical principle behind its function, a
recently discovered surface instability, and demonstrate that the combination of a high
grafting density, an inflated end-group size and a high degree of monodispersity are
conditions for an optimal functionality of the switch.'' In this way, atomistic simulations
of polymer brush involving materials are expected to increase  their predictive power and
help to create a variety of smart, switchable, and multifunctional
surfaces and thin films. Applications of these versatile and multifunctional brush coatings
are envisioned in many areas including fluid control, microfluidics, and thin film sensors
as well as some objects of interest in molecular biology \cite{sommer2}.

\section{Summary}\label{sec8}

To summarize, in this text we have reviewed the application of
DF approaches and  computer simulation methods
for a
description of fluid-tethered polymer brushes interfaces. Several
illustrations of important findings were given. The topic is
very ample and, consequently, we were unable to describe it in full detail.

Albeit the success in application of the theories outlined in this review,
they will benefit of further improvements along different
lines. In particular, a better description of the model of tethered
polymer chains is necessary. Thermodynamic perturbation theory of the
first order involved in the majority of present studies does not permit
to capture structural peculiarities of several systems of interest.
DF approaches encounter difficulties in an
accurate description of attractive interparticle  and
long-range electrostatic interactions. A comparison of theoretical predictions with
computer simulation data is, therefore, highly desirable
to avoid incorrect interpretation of some theoretical results. As for technical
issues necessary to be solved  in the nearest future,
it is of interest to include statistical mechanical analogue of mass
action law into the available theories. This would permit to explore
the effects of chemical association  more in detail.
We are convinced that theoretical
developments in the area would permit to find novel phenomena faster
and with less cost compared to other methods.

Our review also presents  selected
applications of computer simulations in the studies
of systems involving tethered chains. First, we briefly described
some results obtained from DPD coarse-grained simulations
and then we outlined the application of microscopic MC and MD
simulations to explain the mechanism of separation in RPLC.
With an increase in performance of computer facilities, the
role of simulations will increase, and the simulation methods
 in many cases will replace not only theoretical approaches, but also
some experiments. This is particularly true for the cases
of complex systems and for out of equilibrium phenomena, such as
fluid flows.
One of possible directions toward the development of simulation
techniques to describe brush-fluid interfaces would be the application
of multiscale modelling using field-theoretic methodologies~\cite{last1}.
According to those methods, some fragments of polymers are
treated on a fully atomistic (or even {\it ab initio}) level and
the information is used for parametrization of coarse-grained potentials~\cite{last2}.
Such a technique has  already been used to describe the
formation of micelles and vesicles in amphiphilic
systems~\cite{last3}. However, when performing coarse-graining, one
should always keep in mind that the
coarse-grained systems are physically different
from the initial, molecular systems and that the coarse-graining methods cannot
be used as a black box and require a thorough cross-checking.

\section*{Acknowledgements}
This work was  supported by the EU under IRSES Project
STCSCMBS 268498.


\clearpage

\ukrainianpart

\title{Опис інтерфейсів між флюїдом та приєднаними полімерними ланцюжками:
прогрес у застосуванні теорій функціоналу густини та позаґраткових комп'ютерних симуляцій}

\author{С. Соколовський\refaddr{label1}, Я. Ільницький\refaddr{label2},
О. Пізіо\refaddr{label3}}

\addresses{
\addr{label1} Вiддiл моделювання фiзико-хiмiчних процесiв, унiверситет Марiї Кюрi-Склодовської, \\
20031 Люблiн,  Польща
\addr{label2} Інститут фізики   конденсованих систем НАН України, вул. Свєнціцького, 1, 79011, Львів,   Україна
\addr{label3} Інститут хімії, Національний автономний університет м. Мехіко,  Мехіко, Мексика}

\makeukrtitle

\begin{abstract}
\tolerance=3000
Інтерфейси між флюїдом та приєднаними полімерними ланцюжками зустрічаються
у численних наноскопічних об'єктах. Дослідження їх властивостей є цікавим як з точки
зору фундаментальних наукових знань, так і для низки застосувань, зокрема для
проектування нанопристроїв із спеціальним призначенням. В огляді подано
сучасний стан досягнень теоретичних підходів в цій області (насамперед методів,
що базуються на функціоналі густини), які просунули наше розуміння мікроскопічних
властивостей таких інтерфейсів. Ці теоретичні дослідження дозволяють
описати залежності адсорбційних, змочувальних та сольватаційних сил
та особливості електро-інтерфейсних явищ від термодинамічного стану та
характеристик приєднаних ланцюжків. Зроблено також огляд комп'ютерних
симуляцій таких систем, які, як і теоретичні дослідження, співставлено з
наявними експериментальними даними.

\keywords приєднані ланцюжки, полімерні щітки, флюїд,
адсорбція, змочування, розчини електролітів, диференціальна ємність

\end{abstract}

\end{document}